\documentclass[11pt,english]{article}

\renewcommand{\Im}{{\rm Im \, }}
\renewcommand{\Re}{{\rm Re \, }}
\newcommand{\im}{{\rm Im \, }}
\newcommand{\re}{{\rm Re \, }}

\usepackage[latin9]{inputenc}
\usepackage{geometry}
\usepackage[greek,USenglish]{babel}
\usepackage{verbatim}
\usepackage{prettyref}
\usepackage{amsmath}
\usepackage{amssymb}
\usepackage{graphicx}
\usepackage{setspace}
\usepackage{esint}
\usepackage{color}
\definecolor{darkgreen}{rgb}{0,0.5,0}
\definecolor{darkblue}{rgb}{0,0,0.6}
\definecolor{purple}{rgb}{0.4,.2,0.7}
\usepackage[colorlinks=true,citecolor=darkgreen,linkcolor=darkblue,urlcolor=purple]{hyperref}

\numberwithin{equation}{section}
\numberwithin{figure}{section}
\numberwithin{table}{section}

\def\absval#1{\left|#1\right|}

\begin{document}


~
\vskip15mm

\begin{center} {{\huge \textsc{Hot Halos and Galactic Glasses}} \\  \vskip5mm \textsc{ (Carbonado) } }  

\vskip15mm

Dionysios Anninos${^a}$, Tarek Anous$^b$, Jacob Barandes$^b$, Frederik Denef$^{\;b,c}$, Bram Gaasbeek$^c$

\vskip5mm

{
\it{$^a$ Stanford Institute for Theoretical Physics, Stanford University} 

\it{$^b$ Center for the Fundamental Laws of Nature, Harvard University} 

\it{$^c$ Institute for Theoretical Physics, University of Leuven} }

\vskip5mm

{\tt{danninos@stanford.edu, tanous@physics.harvard.edu, jbarandes@physics.harvard.edu, \newline denef@physics.harvard.edu, bram$\_$mmm@hotmail.com}}

\end{center}

\vskip15mm

\begin{abstract}

We initiate a systematic study of the state space of non-extremal, stationary black hole bound states in four-dimensional $\mathcal{N} = 2$ supergravity. Specifically, we show that an exponential multitude of classically stable ``halo'' bound states can be formed between large finite temperature D4-D0 black hole cores and much smaller, arbitrarily charged black holes at the same temperature. We map out in full the regions of existence for thermodynamically stable and metastable bound states in terms of the core's charges and temperature, as well as the region of stability of the core itself. Several features of these systems, such as a macroscopic configurational entropy and exponential relaxation timescales, are similar to those of the extended family of glasses. We draw parallels between the two with a view toward understanding complex systems in fundamental physics. 



\end{abstract}

\newpage

\tableofcontents


\section{Introduction}

\subsection{Hot Halos}

Multicentered black hole and more general multihorizon solutions to Einstein's equations have been an active research area for almost a century \cite{Weyl,majum,pap,Ehlers,Israel,Kinnersley,Bonnor,Kastor:1992nn,SabraLust,Booth:1998gf,Denef:2000nb,Denef:2002ru,Bates:2003vx,Tan:2003jz,Jejjala:2005yu,Griffiths:2006tk,Ishihara:2006iv,Giusto:2007tt,Giusto:2007fx,Ford:2007th,Elvang:2007rd,Rogatko:2007kq,Gaiotto:2007ag,Camps:2008hb,Evslin:2008py,Chng:2008sr,Bena:2009en,AlAlawi:2009qe,Stelea:2009ur,Ferrara:2010cw,Emparan:2010sx,Anninos:2010gh,Stelea:2011jm,Bena:2011zw,SteleaNew}. The first solutions discovered had string-like singularities connecting the holes. 
Others were time dependent 
or required highly fine-tuned exact cancelation of forces allowing arbitrary superpositions. 
Stationary, strictly stable, nonsingular solutions in four dimensions were only discovered more recently \cite{Denef:2000nb,LopesCardoso:2000qm,Bates:2003vx}. These solutions can have an arbitrary number of centers and represent genuine, molecule-like bound states, with equilibrium positions determined by balancing conditions between electrostatic, gravitational and scalar forces. 

Supersymmetry was instrumental in the discovery of these black molecules, but is in no way essential for their existence or stability. Indeed, since the black holes are effectively bound together by potentials with deep minima, deforming the supersymmetric solutions slightly away from extremality, say by throwing some neutral particles into the black holes, will not destroy the bound state. Of course, if one keeps on adding mass to the system, the gravitational pull will eventually overpower all the other forces, causing the black molecule to collapse into a single black hole. Nevertheless, for sufficiently cold systems, one expects a plethora of stationary bound states very much like in the supersymmetric case. One of the goals of this paper is to make this argument more precise, and to map out part of the parameter space where such bound states must exist. 

\begin{figure}
\begin{center}
\includegraphics[height=8cm]{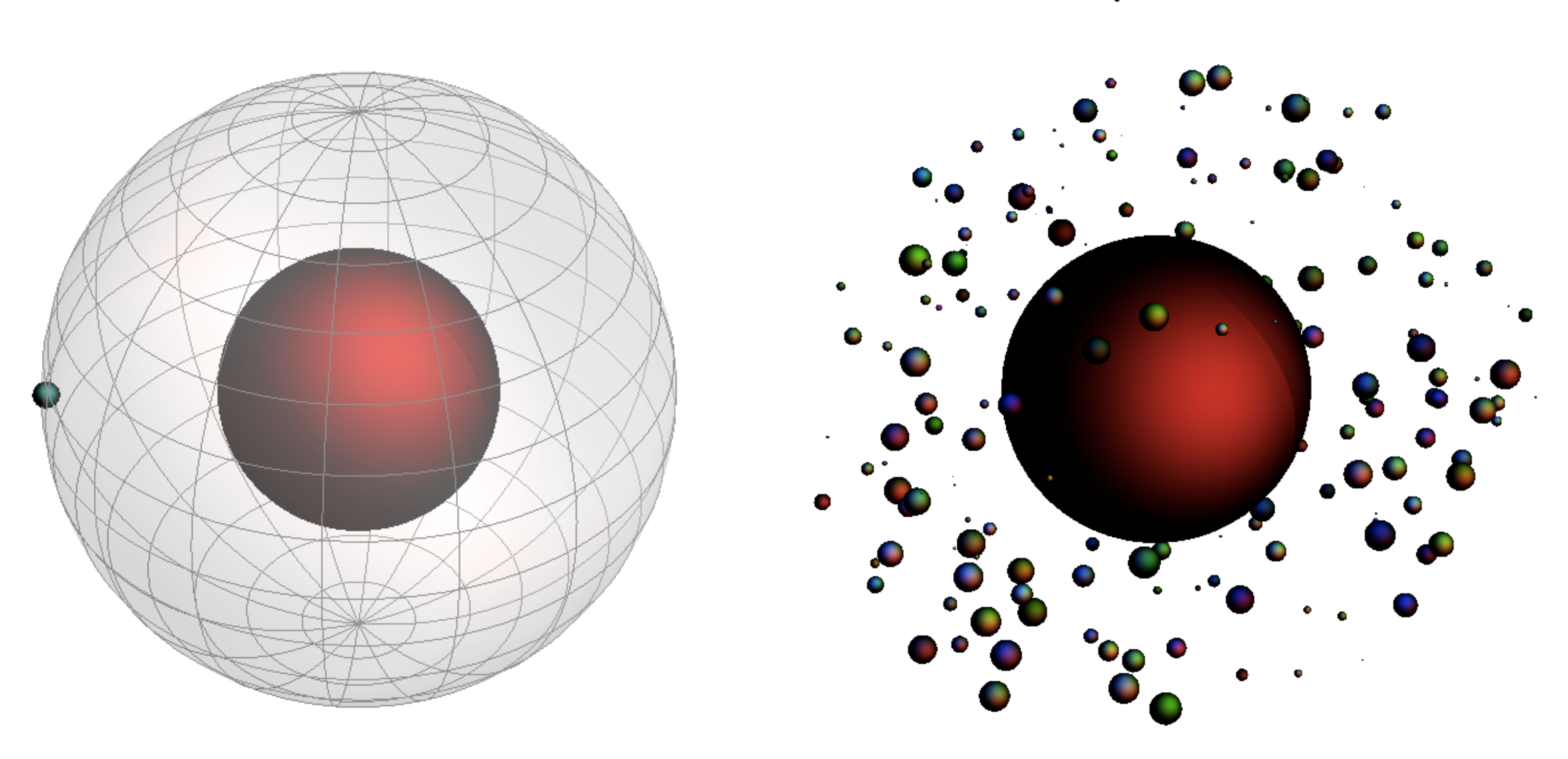}
\end{center}
\vskip-5mm \caption{{\bf Left}: A single probe is allowed to move on a sphere with radius depending on the relative values of its charges. {\bf Right}: Multi-probe bound state configurations consisting of many probe halos orbiting a giant ``galactic'' black hole.  To lowest order in the probe approximation the mutual interactions between the probes are neglected. The lowest energy bound states are stationary: No probes are actually moving, but there is angular momentum stored in the electromagnetic fields, pointing along the radial direction.  Besides spin, the probes induce electric and magnetic dipole charges. In a holographic setup (e.g.\ embedded in AdS$_4$) these would correspond to inhomogeneities in charge densities and magnetic fields in the dual field theory on ${\mathbb R} \times S^2$.
 \label{blackmol}}
\end{figure}

We will do this in the setting of 4d ${\cal N}=2$ supergravity, by considering finite size black holes in the presence of a giant ``galactic'' black hole, all at the same temperature, taking the giant hole large enough so the finite size ones can be treated as probes in a fixed background --- essentially a finite temperature version of the setup of \cite{Andriyash:2010qv}. This is rendered (artistically) in fig.\ \ref{blackmol}. We argue that in this limit the probes can be taken to be BPS without significant loss of generality. Aided by the existence in the theories at hand  of nontrivial families of non-extremal background solutions, an exact computation of the probe interaction potential for any charge is accomplished. It is then a straightforward matter to determine the existence of local minima of this potential and to map out the subset of charges and temperatures where bound states exist and where they are energetically stable. 

We find that many of the interesting features that made their supersymmetric relatives rich and famous, such as wall crossing \cite{Denef:2007vg}, field-induced spin \cite{Denef:2000nb}, intrinsic Landau level degeneracies \cite{Denef:2002ru} and so on, persist at finite temperature. Close to non-BPS extremal limits, we observe that there always exist huge numbers of energetically stable probe bound states, that is to say bound states with free energy less than a single black hole with the same temperature and charge. Close to BPS extremal limits on the other hand, bound states tend to have slightly more energy than the black hole, so although classically stable, they can quantum tunnel into the horizon (though never to spatial infinity). But even on this side we find a region of parameter space where stable bound states exist upon turning on a temperature. 

We hope these results will facilitate searches for exact solutions of nonextremal black hole bound states, or for indirect criteria guaranteeing existence and stability.

\subsection{Galactic Glasses}

Our primary motivation for this work, however, is to further the program of understanding complex systems in string theory in general and particularly, the holography of glass-like systems (other approaches to holographic glasses were proposed in \cite{Adams:2011rj,Kachru:2009xf}). By complex or glass-like systems we roughly mean systems which have free energy landscapes with an exponentially large number of minima. Typically this leads to exponentially slow relaxation, breaking of ergodicity and persistent memory effects (see for example \cite{kurchan}). Examples in the real world include spin glasses \cite{spinglassbook}, i.e.\ spins randomly distributed in a rigid (quenched) environment, effectively giving rise to a disordered mix of ferromagnetic and antiferromagnetic spin-spin interactions. Another and much larger class of examples are structural glasses, like glass itself, or electron glasses \cite{amirreview,amirtalk}, vortex glasses \cite{DuXu} or ice cream \cite{icecream}. Here the disorder does not descend from randomly quenched auxiliary variables, but is generated spontaneously by the system itself, at least when cooled sufficiently fast. Much progress has been made in understanding the nature of glass and the glass transition over the past thirty years, yet it remains one of the deepest unsolved problems in solid state physics \cite{glassproblem,kurchan}. 

Other systems sharing these general glassy features include biological systems such as the brain and other highly complex interacting networks.  In string theory and cosmology, similarly complex systems arise in the description of the landscape, in weak coupling descriptions of black hole microstates obtained from wrapped D-branes, and in the context of eternally inflating spacetimes. Some background material for this was given in \cite{Denef:2011ee,diothesis}. 

The basic observation suggesting black molecules have a holographic interpretation as thermodynamic  states of glassy systems is quite simple. On the one hand, black holes can be thought of as thermodynamic states (in a precise, microscopic way in the context of holography). On the other hand, whenever black molecules exist, there exist exponentially many of them, including highly complex configurations with many centers. In different regimes, macroscopically different configurations can be thermodynamically dominant \cite{Denef:2007vg,deBoer:2008fk,Bena:2011zw}. Thus, a holographic dual description should have microscopic physics leading to a free energy landscape with exponentially many local minima and high complexity. Indeed, already at weak coupling, wrapped D-brane systems do have such energy landscapes \cite{Gomis:2005wc,Denef:2007vg,Denef:2011ee}.

Furthermore, as mentioned earlier, black molecules will eventually collapse when heated up or otherwise brought far from extremality. This process is classical and fast, and can be thought of as the equivalent of glass melting and entering the liquid phase. Going in the other direction however, even if thermodynamically favored, involves exponentially large relaxation times. Starting with a black molecular configuration that has a charge distribution over the centers that is macroscopically different from the minimal free energy configuration, the time scale to get to the equilibrium state may easily become exponentially long, for several reasons. First, the escape rate of a particle from a black hole is proportional to $e^{-\Delta S}$, where $\Delta S$ is the amount of Bekenstein-Hawking entropy the emission costs the black hole \cite{Kraus:1994by,KeskiVakkuri:1996xp,Parikh:1999mf}. Since charge needs to be moved from one black hole center to another to reach equilibrium, $\Delta S$ generically has a gap, due to charge quantization and the BPS bound. Since entropy scales quadratically under uniform rescaling of charge and mass, a generic $\Delta S$ for the emission of a charged particle will scale linearly with the size of the black hole, leading to typical relaxation time scales that are exponential in the system size. Second, even if there happen to be some relatively fast decay channels away from a particular configuration, relaxation is likely to get slowed down tremendously by the existence of an exponentially large landscape of black molecule configurations, much like what happens in actual glasses. Indeed there is in general no reason why the fastest decay channels out of a particular configuration should drive the system straight to the configuration that globally minimizes the free energy. Instead the system may get stuck in deep local minima of the free energy, requiring again exponentially suppressed transitions to get unstuck. 

Finally, it seems plausible that these systems exhibit memory effects similar to those observed in glasses. When a glassy material is perturbed by a constant force for some time $t_p$ and then allowed to relax back for a time $t$, one frequently observes that it approaches its original state as $\delta(t) \sim \log(1+t_p/t)$ over many decades of time \cite{amirreview,amirtalk}. A striking feature of this is its completely universal form and the absence of any intrinsic time scale, replaced instead by a persistent ``memory'' of $t_p$. Such behavior is explained robustly
\cite{amirreview,amirtalk} when one assumes a high dimensional state space with linear stochastic dynamics given by exponentially small transition matrix elements $\Gamma_{ij} \sim e^{-A_{ij}}$, and roughly uniformly (say polynomially) distributed values of $A_{ij}$. The density of transition channels with relaxation time $T=e^A$ is then proportional to $\frac{dT}{T}$ at large $T$, i.e.\ approximately scale invariant. This naturally leads to an approximately scale invariant spectrum of the transition rate matrix, producing the above logarithmic time evolution. Since the systems of interest to us exhibit a similarly broad quasi continuum of exponentially suppressed transition rates, it is reasonable to expect similar memory effects to arise.


Thus it is conceivable that these systems are part of the large family of glassy systems. On the other hand, it is far from obvious if and to what extent they behave like conventional glasses, what their properties are and what their holographic interpretation is, or what the potential lessons are we can draw from their study regarding more general complex systems in string theory. In this paper we will take the necessary first steps to address these questions. 


\subsection{Layout of the paper}

The main focus of our work consists of exploring the nature of finite temperature black hole bound states in $\mathcal{N} = 2$ supergravity in four dimensions. To be more or less self-contained we begin by reviewing in section \ref{setup} the relevant Lagrangians and notations (more background can be found in \cite{bertthesis,jacob}). In section \ref{sec:Single-Center-Non-Extremal-Background} we present a simple consistent truncation scheme valid for any ${\cal N}=2$ string compactification, within which we derive and describe a family of exact non-extremal black hole solutions (previously found in \cite{Galli:2011fq}). These will serve as our galactic black hole cores, dressed by halos of much smaller probe black holes. The probe potentials are introduced in section \ref{probes} and the existence of nonextremal bound states is established. A systematic exploration of the parameter regime in which metastable and stable bound states exist is given in section \ref{sec:analysis}; the results are shown in fig.\ \ref{ST}. We discuss the thermodynamic properties of the system and find a phase structure confirming several of the glass-like features outlined above. The diagram is suggestive of quantum critical points attained by dialing the asymptotic moduli to the black hole attractor fixed point. Across this point, the ratio of probe-induced spin to probe induced D6 magnetic dipole moment changes sign. In the appendix we estimate the configurational entropy of multi-probe BPS galaxies and find it grows linearly with the system's size (charge), and that it scales to zero with a nontrivial exponent near the critical points.

\vskip5mm 
\noindent {\bf Note added}: In interesting recent work \cite{Bena:2011fc,Chowdhury:2011qu}, analogous but complementary finite temperature bound states in five dimensions were independently explored, with qualitatively similar results (see appendix D of \cite{Chowdhury:2011qu} for a detailed comparison).

\section{Setup and notation}\label{setup}

Four dimensional ${\mathcal N}=2$ supergravity coupled to massless vector and hypermultiplets has a bosonic action of the general form \cite{deWit,italians}
\begin{eqnarray}
S_{\mathrm{4D}} 
 &  = & \frac{1}{8\pi}\int d^{4}x \, \sqrt{-g} \left( \tfrac{1}{2} R- G_{A\bar{B}} \, \partial_\mu z^{A} \partial^\mu \bar{z}^{\bar{B}}-h_{XY} \, \partial_\mu q^{X} \partial^\mu q^{Y} \right) \nonumber \\
 &  & +\frac{1}{16\pi}\int d^{4}x \, \sqrt{-g} \left( \im \, \mathcal{N}_{IJ} \, F_{\mu \nu}^{I} F^{J \mu \nu}-\re \,\mathcal{N}_{IJ} \, F^{I}_{\mu \nu}  \tilde{F}^{J \mu \nu} \right) \, ,\nonumber 
\end{eqnarray}
where the $z^A$ $(A=1,\dots,n)$ are the vectormultiplet scalars, $F^{I}_{\mu \nu}=\partial_\mu A_\nu^I-\partial_\nu A_\mu^I$ $(I=0,1,\dots,n)$ are the vector field strengths, $\tilde{F}_{\mu \nu} \equiv \frac{1}{2} \epsilon_{\mu\nu\rho\sigma} \tilde{F}^{\rho \sigma}$, and the $q^{X}$ are the hypermultiplet scalars. We put $G_N = 1$. As $G_{A \bar{B}}$ and ${\cal N}_{IJ}$ only depend on the $z^A$, and $h_{XY}$ depends only on the $q^X$, the hypermultiplets decouple from the vector multiplets and we will not need to consider them further. 
The vectors $A^I$ are sourced by electric charges $Q_I$ and magnetic charges $P^I$.\footnote{Magnetic charges have an upper index, but we will further on often use lower indices for both electric and magnetic charges, to make expressions involving powers of magnetic charges less clumsy.} The space of charge vectors $\Gamma = (P^I,Q_I)$ carries a canonical, duality invariant, symplectic product, which in the standard symplectic basis can be expressed as
\begin{equation}
 \langle \Gamma,\tilde{\Gamma} \rangle = P^I \tilde{Q}_I - Q_I \tilde{P}^I \, .
\end{equation}
The metric $G_{A\bar B}$ is special K\"ahler, {\it i.e.} it is derived from a prepotential $F(X)$:
\begin{equation} \label{GAB}
 G_{A \bar B} = \partial_{z^A} \bar{\partial}_{\bar{z}^{\bar B}} {\mathcal K} \, , \qquad {\mathcal K} = -\log i \langle \Omega,\overline{\Omega} \rangle \, , \qquad \Omega = (X^I,\partial_{X^I} F) \, , \qquad X^A = X^0 z^A \, .
\end{equation}
The variable $X^0$ drops out of all observable quantities; we gauge fix $X^0 \equiv 1$.
The prepotential $F(X)$ is a locally defined holomorphic function, homogeneous of degree 2 in the $X^I$. It also determines the electromagnetic couplings ${\cal N}_{IJ}$:
\begin{equation}\label{NIJ2}
{\cal N}_{IJ} = \bar{F}_{IJ} + 2i \frac{(\Im F_{IK}) X^K (\Im F_{JL})
X^L}{X^M (\Im F_{MN}) X^N}, \qquad F_{IJ} = \partial_{X^I} \partial_{X^J} F \, .
\end{equation}

In type IIA Calabi-Yau compactifications the coordinates $z^A = B^A + i J^A$ are identified complexified K\"ahler moduli and, ignoring string worldsheet instanton corrections, the prepotential takes the form
\begin{equation} \label{cubicF}
 F(X) = \frac{1}{6 X^0} D_{ABC} X^A X^B X^C 
\end{equation}
where the $D_{ABC}$ are triple 4-cycle intersection numbers. The charges $(P_0,P_A,Q_0,Q_A)$ are identified with wrapped (D6,D4,D0,D2) charges. For a Calabi-Yau with a single K\"ahler modulus  $z = z^1$ ({\it i.e.} $n=1$), the cubic prepotential (\ref{cubicF}) becomes
\begin{equation} \label{onemodF}
 F(X) = D \frac{(X^1)^3}{6 X^0} \, . 
\end{equation}
The examples we consider in this paper will all be effectively reducible to this case. The special K\"ahler metric (\ref{GAB}) is then just the Poincar\'e metric on the upper half plane:
\begin{equation} \label{Gzz}
 G_{z \bar{z}} = \frac{3}{4 \, ({\rm Im} \, z)^2} \, ,
\end{equation}
and the electromagnetic coupling matrix (\ref{NIJ2}) becomes, with $z = x+ i y$,
\begin{equation} \label{Nmatrix}
 {\cal N} = D \begin{pmatrix} 
   \frac{x^3}{3}+\frac{i y x^2}{2}+\frac{i y^3}{6} & -\frac{x^2}{2}-\frac{i y x}{2} \\
 -\frac{x^2}{2}-\frac{i y x}{2} & x+\frac{i y}{2}
 \end{pmatrix}
\end{equation}
${\cal N}=2$ supersymmetry implies that the mass of any state of charge $\Gamma$ in a vacuum with asymptotic moduli $z_0$ is bounded below by the absolute value of the central charge $Z(\Gamma,z_0)$, defined by
\begin{equation}  \label{Zgendef}
 Z(\Gamma,z) = - e^{{\mathcal K}/2} \, \langle \Gamma, \Omega \rangle \, .
\end{equation}
States saturating this bound are supersymmetric and called BPS. In the $n=1$ case (\ref{onemodF}), for a charge $\Gamma = (P_0,P_1,Q_0,Q_1)$, (\ref{Zgendef}) becomes 
\begin{equation} \label{centralchargedef}
 Z(\Gamma,z) = \frac{\sqrt{3}}{2 \, \sqrt{D (\Im z)^3}} \left( \tfrac{D}{6} P_0 z^3-\tfrac{D}{2} P_1
   z^2+Q_1 z +Q_0 \right) \, .
\end{equation}
To write down the action of a point particle in a general background, it is convenient to introduce the dual magnetic field strengths 
\begin{equation}
 G_I = \im \, \mathcal{N}_{IJ} \, \tilde{F}^{J}-\re \,\mathcal{N}_{IJ} \, F^{J} \, .
\end{equation} 
The electromagnetic equation of motion $dG_I=0$ implies the existence of dual magnetic gauge potentials $B_I$ such that $G_{I\mu\nu} = \partial_\mu B_{I\mu} - \partial_\nu B_{I \nu}$. Collecting the electric and magnetic gauge field strengths and potentials into duality covariant vectors ${\mathbb F} = (F^I,G_I)$ and ${\mathbb A} = (A^I,B_I)$, the action for a point particle of mass $m$ and charge $\gamma=(p^I,q_I)$ is \cite{Denef:2000nb,Billo:1999ip} 
\begin{equation} \label{probeaction}
 S_\gamma = - \int m \, ds  - \frac{1}{2} \int \langle \gamma, {\mathbb A_\mu} \rangle dx^\mu \, .
\end{equation}
The mass depends on the scalars $z$. In particular when the particle is BPS, we have
\begin{equation}
 m = |Z(\gamma,z)| \, .
\end{equation}

\section{Non-extremal black hole background} \label{sec:Single-Center-Non-Extremal-Background}

We will now construct a class of exact, spherically symmetric, nonextremal single centered black hole solutions for any prepotential of the form (\ref{cubicF}). These are essentially the solutions found in \cite{Galli:2011fq,Gibbons:1982ih}. In the nonsupersymmetric extremal limit they belong to the class studied in \cite{Ceresole:2007wx,Perz:2008kh,Ferrara:2008ap,Gimon:2009gk}.

\subsection{Equations of motion}

The black hole metric is of the general form
\begin{equation} 
 ds^{2}=-e^{2U\left(\tau\right)} \, dt^{2}+e^{-2U\left(\tau\right)} \, \left(\frac{c^{4}}{\sinh^{4}c\tau}d\tau^{2}+\frac{c^{2}}{\sinh^{2}c\tau}d\Omega_{2}^{2}\right),\label{eq:SphSymmLineElemWithouth}
\end{equation}
where $\tau$ is an (inverse) radial coordinate with $\tau=0$ corresponding to spatial infinity and $\tau=\infty$ to the horizon. The parameter $c$ is a positive constant parametrizing the deviation from extremality. 

The scalars depend on $\tau$ only, and the electromagnetic field is given by\footnote{In form notation, $F=\frac{1}{2} F_{\mu \nu} dx^\mu \wedge dx^\nu$, $A = A_\mu dx^\mu$.}
\begin{equation} \label{generalF}
 F^I = P_I \, \omega - (\Im {\cal N})^{IJ} (Q_J + \Re {\cal N}_{JK} P_K) \, \star \omega \, , \qquad \omega = \sin \theta \, d\theta \wedge d\phi \, , \qquad
 \star \omega=e^{2U}d t\wedge d\tau
\end{equation}
Note that this automatically satisfies the Bianchi identity $dF^I=0$, because the ${\cal N}_{IJ}$ and $U$ depend on $\tau$. Moreover, for this particular form of $F^I$, we have $G_I = Q_I \omega + (\cdots)  \star \omega$, hence  the equations of motion $d G_I = 0$ are also automatically satisfied.

The scalar and metric equations of motion can be obtained from an effective particle action \cite{Ferrara:1997tw}
\begin{equation}
S_{\mathrm{eff}}=\int_{0}^{\infty}d\tau\left(\dot{U}^{2}+G_{A \bar{B}}\dot{z}^A\dot{\bar{z}}^{\bar B} - V_{\rm eff}(U,z,\bar{z}) \right) \, ,\label{eq:1DEffActDiagT6}
\end{equation}
with effective potential
\begin{equation}
 V_{\rm eff} = - c^2 - e^{2U} \bigl( \absval{Z}^{2}+4G^{A\bar{B}}\partial_{A}\absval{Z}\bar{\partial}_{\bar{B}}\absval{Z} \bigr) \, , \label{eq:ScalarPotDiagT6}
\end{equation}
supplemented with the constraint that the total particle energy must vanish:
\begin{equation}
\dot{U}^{2}+G_{A\bar{B}}\dot{z}^{A}\dot{\bar{z}}^{\bar{B}} + V_{\rm eff}  = 0 \label{eq:cSqConstraint}~.
\end{equation}

\subsection{Consistent truncations}

Solving this system in general appears intractable, but special classes of solutions can nevertheless be found. First, the general problem with an arbitrary number $n$ of vector multiplets can be consistently reduced to an effective single vector multiplet problem by the  truncation \cite{Denef:2007vg}, for any choice of constant $K^A$ (inside the K\"ahler cone):
\begin{equation}
 z^A = K^A \hat{z}^1 \, , \qquad (F^0,F^A,G_0,G_A) = \left(\hat{F}^0,K^A \hat{F}^1,K^3 \hat{G}_0,(K^2)_A \hat{G}_1 \right) \, ,
\end{equation}
where $(K^2)_A \equiv D_{ABC} K^B K^C$ and $K^3 \equiv D_{ABC} K^A K^B K^C$. It is easily checked that the equations of motion then consistently reduce to the $n=1$ case (\ref{onemodF}), with $D = K^3$. To remain consistent we must also choose the black hole charge to be of the form
\begin{equation}
 \Gamma = \left(\hat{P}^0,K^A \hat{P}^1,K^3 \hat{Q}_0,(K^2)_A \hat{Q}_1 \right) \, ,
\end{equation}
which sources the reduced fields as a charge $(\hat{P}^I,\hat{Q}_I)$ in the effective $n=1$ theory. We will henceforth normalize $K^A$ such that $D = K^3 = 1$, and drop the hats on the reduced quantities.

The effective metric on the scalar space parametrized by $z^1 \equiv z \equiv x+iy$ is given by (\ref{Gzz}) and the effective potential is $V_{\rm eff}(U,x,y)=c^2-e^{2U} V(x,y)$ with 
 \begin{eqnarray}
V = \frac{3}{y^3} \left(Q_{0}+Q_{1}x-\tfrac{1}{2} P_{1}x^{2}+\tfrac{1}{6}P_{0}x^{3}\right)^{2} 
 + \frac{1}{y} \left(Q_{1}-P_{1}x+\tfrac{1}{2} P_{0}x^{2}\right)^{2}
 +\frac{y}{4} \left(P_{1}-P_{0}x\right)^{2}+ \frac{y^{3}}{12} P_{0}^2 .\label{eq:ScalarPotDiagT6QP}
\end{eqnarray}
The resulting equations of motion are still hard to solve in the generic case, but when $x=0$, the coupling matrix (\ref{Nmatrix}) becomes pure imaginary and diagonal, and the system simplifies considerably. This motivates a search for solutions with constant $x(\tau)=0$. In this case the fields strengths (\ref{generalF}) are of the simple form
\begin{equation} \label{FGexpr}
 \begin{array}{ll}
  F^0 = P_0 \,\omega - Q_0  \frac{6}{y^3} \star \omega\, , \qquad &
  F^1 = P_1 \,\omega - Q_1 \frac{2}{y} \star \omega \, , \\
  G_0 = Q_0 \,\omega + P_0 \frac{y^3}{6} \star \omega \, , \qquad &
  G_1 = Q_1 \,\omega + P_1 \frac{y}{2} \star \omega \, .
 \end{array}
\end{equation} 
Consistency of the ansatz $x(\tau)=0$ requires $\partial_x V|_{x=0} = 0$, which leads to the conditions $Q_0 Q_1=Q_1 P_1=P_0 P_1=0$, leaving the possibility to have D4-D0 ($\Gamma=\left(0,P_{1},Q_{0},0\right)$),
D6-D2 ($\Gamma=\left(P_{0},0,0,Q_{1}\right)$), or D6-D0 ($\Gamma = \left(P_{0},0,Q_{0},0\right)$)
background charges. In the following we specialize to these cases.

\subsection{Solving the equations of motion}

Putting $y = e^\phi$, the remaining equations of motion for $U$ and $\phi$ derived from (\ref{eq:1DEffActDiagT6}) take the form 
\begin{eqnarray}
\ddot{U} &  = & e^{2U}\bigl(v_{1}e^{a\phi}+v_{2}e^{b\phi}\bigr),\label{eq:UeomTrunc}\\
\ddot{\phi} &  = & \tfrac{2}{3} e^{2U}\bigl(av_{1}e^{a\phi}+bv_{2}e^{b\phi}\bigr) \, ,\label{eq:phieomTrunc}
\end{eqnarray}
together with the constraint \eqref{eq:cSqConstraint} (which just fixes the value of $c$). For the D4-D0 system, we have $(v_{1},v_2;a,b)=\left(3Q_{0}^{2},\frac{P_{1}^{2}}{4};-3,1\right)$, for D6-D2 $(v_{1},v_2;a,b)=\left(Q_{1}^{2},\frac{P_{0}^{2}}{12};-1,3\right)$ and for D6-D0 $(v_{1},v_2;a,b)=\left(3Q_{0}^{2},\frac{P_{0}^{2}}{12};-3,3\right)$. This system is of Toda form \cite{Olive:1980sd}. Following the method of \cite{Gibbons:1982ih}, we set $\alpha\equiv2U+a\phi$, $\beta\equiv2U+b\phi$. The system of equations of motion for $U$ and $\phi$ then becomes
\begin{eqnarray*}
\ddot{\alpha} &  = & \alpha_0 \, e^{\alpha}+\gamma_0 \, e^{\beta},\\
\ddot{\beta} &  = & \delta_0 \, e^{\alpha}+\beta_0 \, e^{\beta}.
\end{eqnarray*}
where $\alpha_0=\frac{2}{3}\left(3+a^{2}\right) v_1$, $\beta_0=\frac{2}{3}\left(3+b^{2}\right) v_2$, $\gamma_0=\frac{2}{3}\left(3+ab\right) v_2$, and $\delta_0=\frac{2}{3}\left(3+ab\right) v_2$. These two equations decouple if $ab=-3$, which happens to be the case for the D4-D0 and D6-D2 systems.  In these cases the equations of motion integrate to
\begin{equation}
\begin{array}{rcl}
\alpha\left(\tau\right) &  = & {\displaystyle \log\left(\frac{2c_{1}^{2}}{\alpha_{0}\sinh^{2}\left(c_{1}\tau+c_{2}\right)}\right)},\\
\beta\left(\tau\right) &  = & {\displaystyle \log\left(\frac{2c_{3}^{2}}{\beta_{0}\sinh^{2}\left(c_{3}\tau+c_{4}\right)}\right)},
\end{array}\label{eq:alphabetaFromcs}
\end{equation}
 where $c_{1},c_{2},c_{3},c_{4}$ are positive integration constants and $(\alpha_0,\beta_0)=\left( 24Q_{0}^{2},\frac{2P_{1}^{2}}{3} \right)$ for the D4-D0 while $(\alpha_0,\beta_0)=\left( \frac{8Q_{1}^{2}}{3},\frac{2P_{0}^{2}}{3}\right)$ for the D6-D2.
 
Specializing to the D4-D0 case, this implies for the original fields
\begin{eqnarray}
e^{-4U} &  = & { \displaystyle \frac{2|{Q_{0}P_{1}^{3}}|}{3{c_{1}c_{3}^{3}}} \sinh\left(c_{1}\tau+c_{2}\right)\sinh^{3}\left(c_{3}\tau+c_{4}\right)}  ,\\
y^{2}=e^{2\phi} &  = & {\displaystyle {\frac{6|Q_{0}|}{|P_{1}|}}{\frac{c_{3}}{c_{1}}}\frac{\sinh\left(c_{1}\tau+c_{2}\right)}{\sinh\left(c_{3}\tau+c_{4}\right)}} .
\label{eq:UphiFromcs}
\end{eqnarray}
The constraint (\ref{eq:cSqConstraint}) fixes $c^2=(c_1^2+3c_3^2)/4$. Regularity of $\phi$ at the horizon $\tau=\infty$ requires $c_1=c_3$. The asymptotic boundary conditions $U(\tau=0)=0$, $y(\tau=0)=y_0$ further imply
\begin{equation}
 \sinh c_2 = \frac{c \, y_0^{3/2}}{2\sqrt{3} \, |Q_0|} \, ,
\qquad \sinh c_4 = \frac{\sqrt{3} \, c}{|P_1|  \, y_0^{1/2}} \, .
\end{equation}

\subsection{The D4-D0 solution}

Putting everything together, denoting\footnote{This notation is motivated by the fact that in the $c \to 0$ extremal limit, $H_0$ and $H_1$ become the flat space D0 resp.\ D4 harmonic functions ubiquitous in the description of the well-known extremal solutions.}
\begin{equation}
 H_0 \equiv \frac{|Q_0|}{c} \sinh(c \tau + c_2) \, , \qquad 
 H_1 \equiv \frac{|P_1|}{c} \sinh(c \tau + c_4) \, ,
\end{equation}
we get for the metric warp factor and the scalar
\begin{eqnarray}
e^{-2U} & = & \sqrt{\frac{2}{3} H_0 H_1^3} \\
y& = & \sqrt{\frac{6 H_0}{H_1}} \, ,
\label{eq:UphiFromcs2}
\end{eqnarray}
and for the gauge potentials ${\mathbb A}=(A^I,B_I)$, obtained by integrating the field strengths (\ref{FGexpr}):
\begin{equation} \label{ABexpr}
 \begin{array}{ll}
  A^0 = \frac{1}{2 Q_0} \left( \sqrt{c^2 + \frac{Q_0^2}{H_0^2}} - c\right) \, dt \, , \qquad &
  A^1 = P_1 (1-\cos \theta) \, d\phi \, , \\
  B_0 = Q_0 (1-\cos \theta) \, d\phi  \, , \qquad &
  B_1 = -\frac{3}{2 P_1} \left( \sqrt{c^2 + \frac{P_1^2}{H_1^2}} - c \right) \, dt \, .
 \end{array}
\end{equation} 
We have chosen a gauge here in which the electric potentials vanish at the horizon and the Dirac monopole potentials are regular on the northern sphere.

Notice that the modulus $y$ at the horizon $\tau=\infty$ is fixed at an attractor point $y_\star$ independent of $y_0$ and $c$:
\begin{equation} \label{attractorpoint}
 y_\star = \sqrt{\frac{6 |Q_0|}{|P_1|}} \, .
\end{equation}


\subsection{Mass, entropy, temperature and specific heat} \label{sec:MET}

The ADM mass of the black hole can be read off from the asymptotic behavior of the metric: 
\begin{eqnarray}
 M &=& \frac{c}{4}\left( \coth c_2 + 3 \coth c_4 \right) \\
 &=& \frac{1}{4} \sqrt{c^2 + \frac{12 \, Q_0^2}{y_0^3}}+
 \frac{3}{4} \sqrt{c^2 + \frac{P_1^2 \, y_0}{3}}  \, . \label{MADM}
\end{eqnarray}
In the extremal limit $c = 0$, this becomes $M=\frac{\sqrt{3}}{4} |P_1| y_0^{1/2} + \frac{\sqrt{3}}{2} \frac{|Q_0|}{y_0^{3/2}}$. When $P_1$ and $Q_0$ have the same sign, this equals the absolute value of the central charge and the extremal limit is supersymmetric. When $P_1$ and $Q_0$ have opposite sign the mass is strictly larger than the absolute value of the central charge and the extremal limit is nonsupersymmetric. 

The Bekenstein-Hawking entropy can be read off from the near horizon ($\tau \to \infty$) behavior of the metric. Introducing $r \equiv e^{-c \tau}$, this is:
\begin{equation}
 ds^2 \simeq - \frac{4 \pi c^2}{S} r^2 dt^2 + \frac{4 S}{\pi} dr^2 + \frac{S}{\pi} \, d\Omega_2^2 \, ,
\end{equation}
where $S=A_{\rm hor}/4$ is the Bekenstein-Hawking entropy: 
\begin{eqnarray}
 S &=& \pi \sqrt{\frac{2 |Q_0P_1^3|}{3}} \, e^{(c_2+3c_4)/2} \, \\
 &=& \pi \left( c + \sqrt{c^2 + \frac{12 \, Q_0^2}{y_0^3}} \right)^{1/2} \left( c + \sqrt{c^2 + \frac{P_1^2 \, y_0}{3}} \right)^{3/2}  \, .
\end{eqnarray}

The temperature can be read off from the near-horizon metric by Wick rotating the time coordinate and fixing its periodicity $\beta=1/T$ by requiring regularity at the origin $r=0$. This yields
\begin{equation}
 T = \frac{c}{2 S} \, .
\end{equation}

\begin{figure}
\begin{center}
\includegraphics[height=8cm]{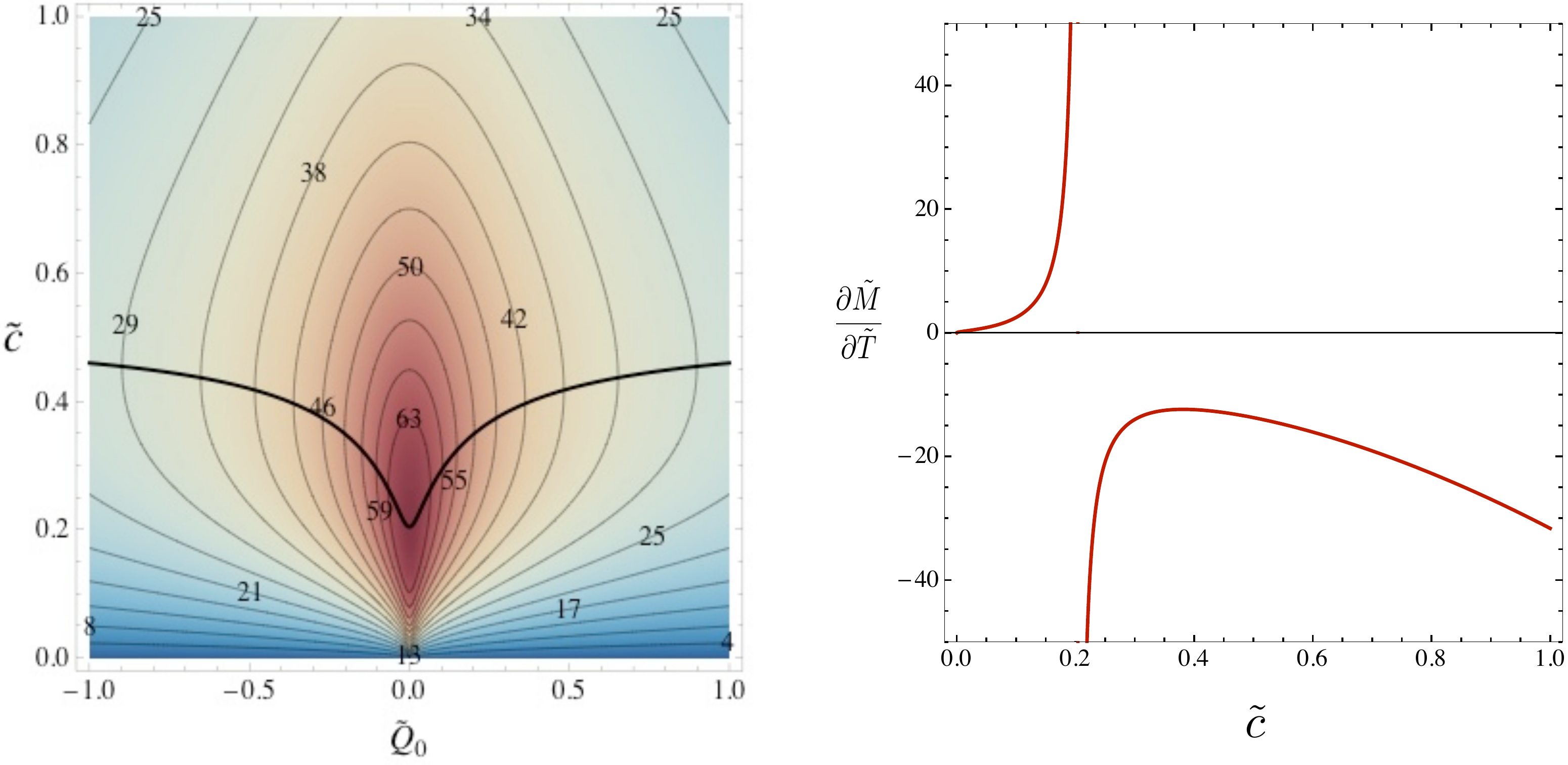}
\end{center}
\vskip-5mm \caption{{\bf (a)}: Rescaled temperature $\tilde{T} \equiv  \sqrt{y_0} P_1 \, T$ as a function of $\tilde{Q}_0$ and $\tilde{c}$, multiplied by $10^3$. Red is warm, blue is cold. The temperature reaches its maximum at fixed $\tilde{Q}_0$ on the thick curve. {\bf (b)}: Specific heat for $\tilde{Q}_0=0$, diverging at $c_{\rm crit} = 1/{2\sqrt{6}} \approx 0.2$, where the temperature reaches a maximum $\tilde{T}_{\rm max} = \sqrt{3}/{8 \pi} \approx 69 \times 10^{-3}$.  
 \label{NV2}}
\end{figure}

Whereas the mass and entropy of the single centered D4-D0 black hole are monotonic function of $c$, this is not so for the temperature, as is clear from figure \ref{NV2} (the rescaled tilde variables will be defined in (\ref{tildevars}) and below): The temperature starts at zero in the extremal limit $c=0$, acquires a maximal value at some intermediate $c$, and goes to zero again in the Schwarzchild black hole limit $c \to \infty$.  The rescaled specific heat $\frac{\partial M}{\partial T}$ at fixed $P_1,Q_0,y_0$ diverges at this point \cite{Davies:1978mf} and changes sign from positive to negative.

\subsection{Scaling symmetries} \label{scalingsymm}

It turns out to be very useful to keep in mind the scaling symmetries $X \to \lambda_1^{n_1} \lambda_2^{n_2} X$, $\lambda_i \in {\mathbb R}^+$, acting on the various quantities defined so far with exponents
\begin{center}
\begin{tabular}{|c|ccccc|cc|ccc|} 
 \hline 
  & $P_1$ & $Q_0$ & $c$ & $y_0$ &$\tau$ & $H_0$ & $H_1$ & $M$ & $S$ & $T$  \\
  \hline
 $n_1$  & 1  & 1 &  1 & 0 & $-1$ &  0 & 0 & 1 & 2 & $-1$  \\
 $n_2$  & 1  & 3 &  $\frac{3}{2}$ & 1 & $-\frac{3}{2}$ &  $\frac{3}{2}$ & $-\frac{1}{2}$ & $\frac{3}{2}$ & 3 & $-\frac{3}{2}$  \\
    \hline
\end{tabular}
\end{center}
The first symmetry descends from a general symmetry of Einstein gravity coupled to vectors, the second from a general symmetry valid for cubic prepotentials \cite{Denef:2007vg}. The scalings of the derived quantities $H_0$, $H_1$, $M$, $S$ and $T$ follow from the scalings of the charges and of $c$, $y_0$ and $\tau$. A consequence of these symmetries is that physical quantities will depend only on invariant combinations of the parameters, up to an overall factor determined by the scaling properties of the quantity under consideration. We choose our independent invariant parameters to be  
\begin{equation} \label{tildevars}
 \tilde{Q}_0 \equiv \frac{Q_0}{y_0^2 P_1} \, \qquad \tilde{c} \equiv \frac{c}{\sqrt{y_0} P_1} \, . 
\end{equation}
A quantity $X$ with scaling exponents $(n_1,n_2)$ will then have the functional dependence 
\begin{equation} \label{scalingrules}
 X(P_1,y_0,Q_0,c) = P_1^{n_1} y_0^{n_2-n_1} \tilde{X}(\tilde{Q}_0,\tilde{c}) \, , \qquad  \tilde{X}(\tilde{Q}_0,\tilde{c}) = X(1,1,\tilde{Q}_0,\tilde{c}) \, .
\end{equation} 
For example the ADM mass (\ref{MADM}) can be written as $M= P_1 \sqrt{y_0} \tilde{M}$ with $\tilde{M} = \bigl( \frac{1}{4} \sqrt{\tilde{c}^2 + 12 \, \tilde{Q}_0^2}+ \frac{3}{4} \sqrt{\tilde{c}^2 + \frac{1}{3}} \bigr)$, the entropy as $S = P_1^2 y_0 \tilde{S}$ and the temperature as $T = \tilde{T}/P_1 \sqrt{y_0}$.




\section{Probe bound states}\label{probes}

In this section we explicitly demonstrate the existence of bound states of probe particles in the nonextremal D4-D0 black hole backgrounds described in the previous section. 

\subsection{BPS probes}

We will in this paper primarily consider probe particles of charge $\gamma=(p_0,p_1,q_0,q_1)$ that are themselves BPS. These ``particles'' could themselves be large black holes, but they must be much smaller than the background black hole so backreaction can be neglected. In thermal equilibrium the probe black hole will acquire the temperature of the background black hole, so it will not quite be BPS. However, the background temperature is parametrically suppressed as $T \sim 1/M$ in the limit of large background black hole mass $M$, as can be seen explicitly from the scaling table in section \ref{scalingsymm}. Hence for any fixed probe size, the thermal contribution to the probe energy, which is proportional to $T$, will vanish in the large $M$ limit, and thus the BPS approximation is justified.\footnote{More precisely, the probe thermal energy is of order $E_T \sim T S_p$, with $S_p$ the probe entropy. Scaling up the background charges uniformly by a factor of $\lambda_1$ while keeping the probe fixed scales $E_T \propto T \propto 1/\lambda_1 \to 0$, whereas the probe potential remains invariant (as we will confirm below). Assuming this potential is not exactly flat, we can therefore neglect the thermal energy. Strictly speaking, this argument only tells us we can assume the probe to be extremal, but not necessarily BPS. However non-BPS extremal black holes are expected to be unstable to decay into lighter particles, as will be confirmed explicitly in section \ref{sec:analysis}, and this on a time scale exponentially smaller than possible instabilities of the background black hole. This justifies considering primarily BPS probes. \label{nothermal}}   

The static potential for a BPS particle can be obtained from (\ref{probeaction}) and the solutions found in the previous section. It consists of two parts, a gravitational part $V_{\rm g}=e^U |Z(\gamma,y)|$ and an electromagnetic part $V_{\rm em} = \frac{1}{2} \langle \gamma, {\mathbb A}_0 \rangle$. Explicitly $V_p=V_{\rm g} + V_{\rm em}$ with
\begin{equation} \label{probepotdef}
 V_{\rm g} = \frac{1}{4} \sqrt{ 
  \left( \frac{q_0}{H_0} + \frac{3 \, p_1}{H_1} \right)^2 + \frac{6 \, H_0}{H_1} \left( \frac{q_1}{H_0} - \frac{p_0}{H_1} \right)^2
 } \, .
\end{equation}
and
\begin{equation}
 V_{\rm em} = -\frac{1}{4} \frac{q_0}{Q_0} \left( \sqrt{c^2 + \frac{Q_0^2}{H_0^2}} - c\right) - \frac{3}{4} \frac{p_1}{P_1} \left( \sqrt{c^2 + \frac{P_1^2}{H_1^2}} - c\right) \, , 
\end{equation}

To avoid complications with marginal stability decays of the probe as it moves around in the nontrivial background, we will only consider probes that are themselves single centered black holes or particles. The BPS probe entropy is given by $S_p \equiv \pi \sqrt{{\cal D}}$, where the so-called discriminant $\cal D$ must be positive for a solution to exist; in the case at hand this is \cite{Shmakova:1996nz}: 
\begin{equation} \label{disconstr}
 {\cal D} = \frac{2}{3} \, p_1^3 q_0 - \, p_0^2 q_0^2 - 2 \, p_0 p_1 q_0 q_1 + \frac{1}{3} \, p_1^2 q_1^2 - \frac{8}{9} \, p_0 q_1^3 \geq 0. 
\end{equation}
Finding an explicit parametrization of this subset of charges seems hard, but is actually made easy by using the invariance of ${\cal D}$ under shifts $(p_0,p_1,q_0,q_1) = \gamma \to \gamma_a=(p_0,p_1+ p_0 a,q_0 - q_1 a - p_1 \frac{a^2}{2} - p_0 \frac{a^3}{6},q_1+p_1 a + p_0 \frac{a^2}{2})$ with $a \in \mathbb{R}$. This invariance follows from the axionic shift symmetry $z \to z-a$ of the supergravity theory under consideration, which leaves in particular the black hole entropy invariant.\footnote{Explicitly: ${\cal D}(\gamma) = \min_z |Z(\gamma,z)|^2 = \min_z |Z(\gamma,z-a)|^2 = \min_z |Z(\gamma_a,z)|^2 = {\cal D}(\gamma_a)$.}

Thus, in the parametrization
\begin{equation} \label{monpar}
 p_1/p_0 = k \, , \qquad 
 q_1/p_0 = -b + \frac{k^2}{2}  \, , \qquad
 q_0/p_0 = n + b k -\frac{k^3}{6} \, ,
\end{equation}
we get simply
\begin{equation}
 {\cal D} = p_0^4 \bigl( \tfrac{8}{9} b^3 - n^2 \bigr) \, ,
\end{equation}
and we may explicitly parametrize the solutions to the constraint (\ref{disconstr}) as $b=\left( \frac{9}{8} (n^2 + \frac{{\cal D}}{p_0^4}) \right)^{1/3}$, ${\cal D} \geq 0$. 

In type IIA compactifications, $k$ may be thought of as the $U(1)$ flux on the wrapped D6, which carries no entropy, while $b$ and $n$ are the ``entropic'' contributions to the charges \cite{Denef:2007vg}.

\subsection{Scalings and validity of probe approximation} \label{sec:allscalings}

Besides the scaling symmetries described in section \ref{scalingsymm}, we have an additional  symmetry uniformly scaling only the probe charge $\gamma$, present because we work to linear order in $\gamma$. All in all we get the following scalings $X \to \lambda_1^{n_1} \lambda_2^{n_2} \lambda_3^{n_3} X$:
\begin{center}
\begin{tabular}{|c|ccccc|cccccccccc|} 
 \hline 
  & $P_1$ & $Q_0$ & $c$ & $y_0$ &$\tau$ & $p_0$ & $p_1$ & $q_1$ & $q_0$ & $k$ & $b$ 
 & $n$ &  
  $m_p$ & $S_p$ & $V_p$  \\
  \hline
 $n_1$  & 1  & 1 &  1 & 0 & $-1$ &   0 & 0 & 0 & 0 & 0 & 0 & 0 & 0 & 0 & 0 \\
 $n_2$  & 1  & 3 &  $\frac{3}{2}$ & 1 & $-\frac{3}{2}$ &  0 & 1 & 2 & 3  & 1 & 2 & 3 & $\frac{3}{2}$ & 3 & $\frac{3}{2}$ \\
 $n_3$  & 0  & 0  & 0 & 0 &  0 & 1 & 1 & 1 & 1 & 0 & 0 & 0 & 1 & 2 & 1\\
    \hline
\end{tabular}
\end{center}
The action on the background is the same as before, and we used the third scaling to set to zero the action of the $\lambda_1$-scaling symmetry on the probe charge. In addition to these continuous symmetries, there is a ${\mathbb Z}_2$ symmetry inverting the signs of all charges, and another ${\mathbb Z}_2$ inverting only the signs of D2 and D6-charges. We point out here that the probe potential does not scale with the size of the black hole.

Analogous to (\ref{tildevars}) and (\ref{scalingrules}), we can scale out powers $p_0$ in addition to $P_1$, $y_0$ according to the following scaling dimensions:
\begin{equation}
 X = P_1^{n_1} y_0^{n_2-n_1} p_0^{n_3} \tilde{X} \, .
\end{equation}
So for example $k=y_0 \tilde{k}$, $b = y_0^2 \tilde{b}$, $n= y_0^3 \tilde{n}$, and $S_p = y_0^3 p_0^2 \tilde{S}_p$ with $\tilde{S}_p = \pi \sqrt{\frac{8}{9} \tilde{b}^3 - \tilde{n}^2}$.

Some care has to be taken not to forget the regime of validity of the probe approximation. The ratio probe mass $m_p$ over background black hole mass $M$ has scaling weights $(-1,0,1)$, so $\frac{m_p}{M} = \frac{y_0 p_0}{P_1} \frac{\tilde{m}_p}{\tilde{M}}$. From this we see that for typical tilde variables of order 1 (and $p_0 \neq 0$), we must keep $y_0 p_0 \ll P_1$ to guarantee $m_p/M \ll 1$. Indeed, when $y_0 \sim P_1$, a single pure D6 becomes as massive as $P_1$ D4-branes, i.e.\ the background black hole, thus spoiling the probe approximation. In particular, since $y_0 \to \infty$ is the M-theory decoupling limit \cite{deBoer:2008fk}, this means that black hole bound states in AdS$_3 \times S^2$ (of which the exact supersymmetric versions were constructed in \cite{deBoer:2008fk}) are not reliably captured by the 4d probe analysis of this paper, in particular not for non-BPS configurations. On the other hand, for any fixed value of $y_0$ and the probe charges, we can always send $P_1 \to \infty$ to make the probe approximation arbitrarily accurate.

\subsection{Bound states} \label{sec:bs}

\begin{figure}
\begin{center}
\includegraphics[height=6.9cm]{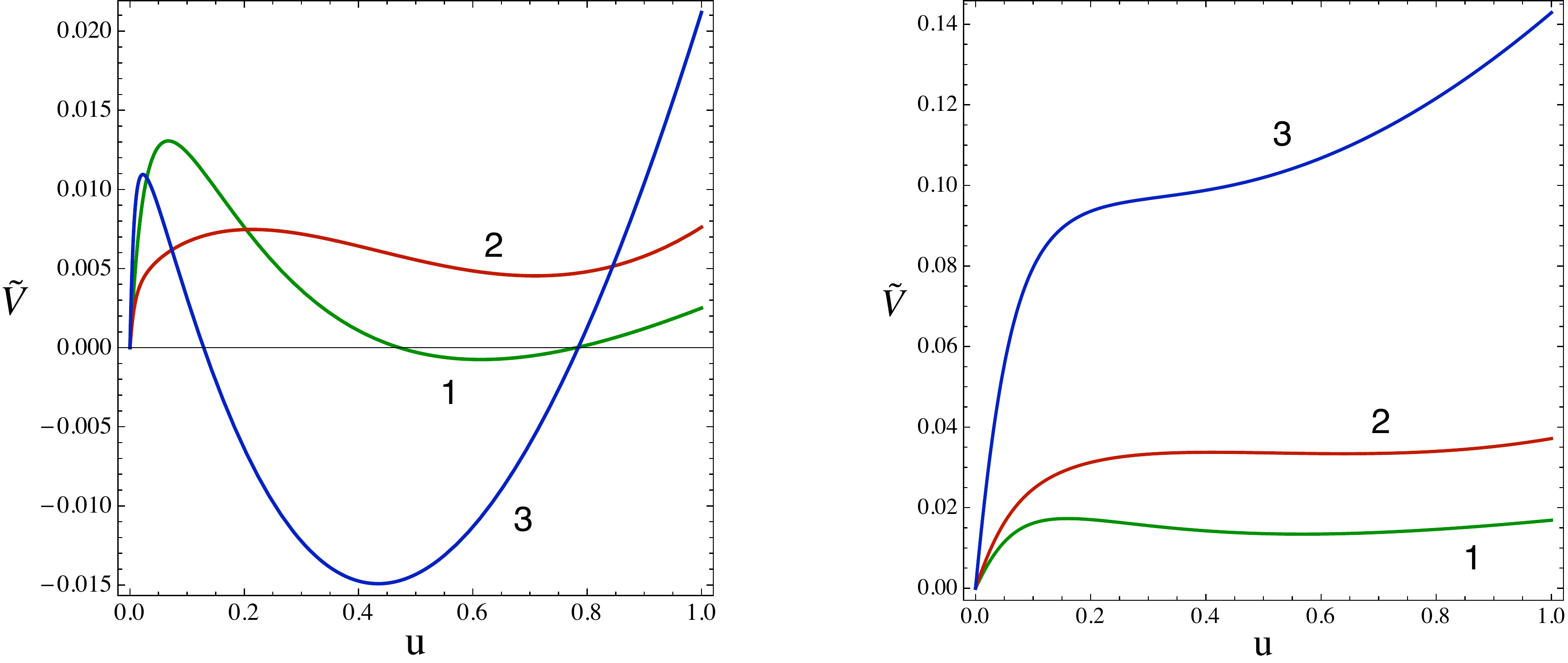}
\end{center}
\vskip-5mm \caption{Examples of probe potentials. Local minima give rise to bound states. The radial $u$ coordinate is defined by $\frac{u}{1-u} = \frac{\tilde{c}}{\sinh(\tilde{c} \tilde{\tau})}$. The horizon is at $u=0$ and spatial infinity at $u=1$. In general, the potential always goes up from the horizon (for $c>0$) and has at most one bump and at most one interior local minimum. On the left we show probe potentials at $(\tilde{Q}_0,\tilde{c},\tilde{k},9\tilde{\cal D},\tilde{n})$ equal to $(.01,.01,.5,0,0)$ for $(1)$, $(.4,.01,.65,0,0)$ for $(2)$ and $(-.4,.01,2,.01,.15)$ for $(3)$.  On the right we show the same but at $\tilde{c} = .08$. At this higher temperature, the minima of $(1)$ and $(2)$ have become positive and more shallow, while $(3)$ has lost its minimum altogether. Increasing $\tilde{c}$ even more wipes out all local minima. 
 \label{potentials}}
\end{figure}

The probe will form a stationary ``molecular'' bound state with the black hole whenever the potential has a nontrivial local minimum. In the supersymmetric case, the discovery of such probe bound states led the way to the construction of general nonlinear black hole bound state solutions in ${\cal N}=2$ supergravity \cite{Denef:2000nb}. In particular their existence made it clear that such bound states had to exist, and quite remarkably, the simple explicit formula for the equilibrium radius obtained from the probe analysis is formally exactly the same as the corresponding formula obtained from supergravity. This was later explained as being a consequence of the constraints imposed by supersymmetry \cite{Denef:2002ru}. There is no reason to expect a similar exact match in the nonextremal case, but a probe analysis will still provide reliable information about the existence of bound states in suitable regimes.

To reproduce first the supersymmetric result, we consider the case of supersymmetric background, $c \to 0$, $P_1,Q_0>0$. 
The probe potential is then of the form $V_p=\sqrt{V_{\rm em}^2+\Delta^2} + V_{\rm em}$. This makes the BPS bound $V_p \geq 0$ manifest. If a BPS-saturating supersymmetric minimum $V_p=0$ exists, it is reached at the radius $r_{\rm eq} = 1/\tau_{\rm eq}$ for which $\Delta(\tau_{\rm eq})=0$ and $V_{\rm em}(\tau_{\rm eq})<0$. The first condition is $q_1 H_1(\tau_{\rm eq}) - p_0 H_0(\tau_{\rm eq}) = 0$, or explicitly, using $\lim_{c \to 0} H_0=Q_0 \tau+\frac{y_0^{3/2}}{2\sqrt{3}}$ and $\lim_{c \to 0} H_1 = P_1 \tau + \frac{\sqrt{3}}{y_0^{1/2}}$: 
\begin{equation} \label{reqformula}
 r_{\rm eq, \, BPS} = \frac{p_0 Q_0 - q_1 P_1}{q_1 \sqrt{\frac{3}{y_0}} - p_0 \sqrt{\frac{y_0^3}{12}} } \, .
\end{equation}
This reproduces the standard BPS equilibrium separation formula for two-centered bound states of this kind \cite{Denef:2000nb}. 



In the nonextremal case there is no such simple expression for $r_{\rm eq}$, but by continuity there will obviously still exist bound states for suitable values of the charges and the nonextremality parameter $c$. Some examples of probe potentials with local minima are shown in fig.\ \ref{potentials}. As suggested by the figure, increasing the nonextremality parameter $c$ typically tends to push up the local minimum, until it eventually disappears altogether and the probe rolls into the black hole. This is to be expected, since going away from extremality means adding more mass. Thus the gravitational pull becomes increasingly more important, eventually overpowering all other forces. In some cases however, in particular for small positive values of $\tilde{Q}_0$, going away from $c=0$ initially \emph{decreases} the value of the potential at the local minimum. An example is potential (1) in the figure (as opposed to (2)): At $c=0$ this is a supersymmetric $(i.e.\ V=0)$ local minimum, whereas for small but nonzero $\tilde{c}$ it is negative. For larger $\tilde{c}$ is goes positive again. Thus, interestingly, slightly heating a supersymmetric black hole with small $\tilde{Q}_0$ will make it unstable to emission of such charges. We view this as an interesting interplay between supersymmetric and thermal physics. For non-BPS extremal black holes $(\tilde{Q}_0<0)$, negative energy probe bound states exist for sufficiently small $\tilde{c}$ for all values of $\tilde{Q}_0$.  An example is potential (3) in the figure.

In the following section we discuss existence and stability in more detail.

\section{Existence, stability and phases} \label{sec:analysis}

\begin{figure}[h]
\begin{center}
\includegraphics[height=7.5cm]{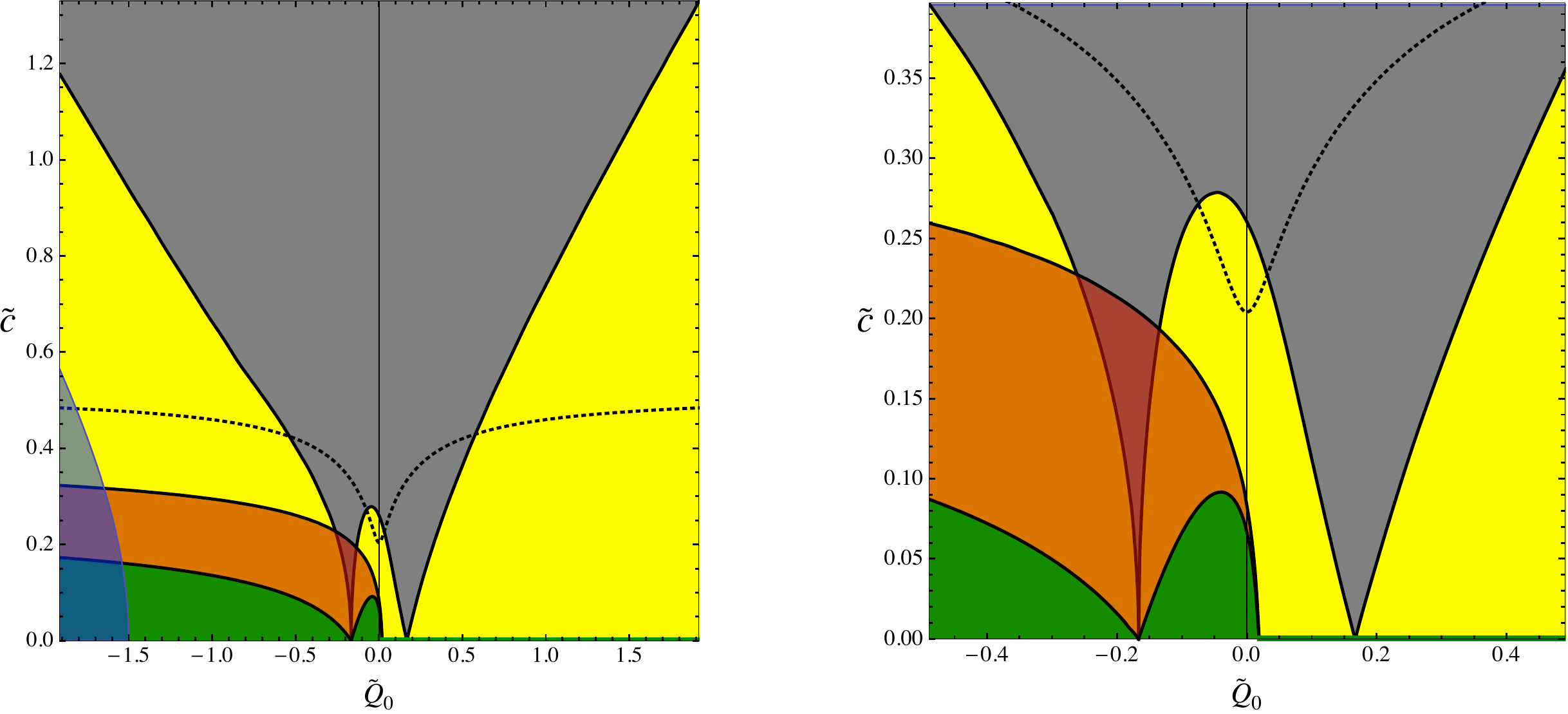}
\end{center}
\vskip-5mm \caption{Existence and stability regions of hot black molecules in the $(\tilde{Q}_0,\tilde{c})$-plane. The figure on the right zooms in on a smaller region but is otherwise the same as the figure on the left. No bound states exist in the grey region. In the yellow region bound states exists, but they are all positive energy (curve (2) in fig. \ref{potentials}). In the green region, negative energy bound states exist (curve (1) in fig.\ \ref{potentials}a). The orange overlay is the region where the black hole core itself is unstable for emission of particles to infinity. The blue overlay in the lower left corner is the region within the $\tilde{Q}_0 < -\frac{1}{6}$ range where pure fluxed D6 probe bound states exist (when $\tilde{Q}_0 > -\frac{1}{6}$ these always exist and moreover they always produce the lowest energy bound states). The dotted line is the maximal temperature line also shown in fig.\ \ref{NV2}. 
The grey regions touch the zero temperature axis at $\tilde{Q}_0=\pm \frac{1}{6}$, or equivalently when $y_0$ coincides with the black hole fixed point $y_\star$. 
 \label{ST}}
\end{figure}

\subsection{Supersymmetric bound states} \label{sec:susyBS}

In the supersymmetric case $\tilde{Q}_0 > 0$, $\tilde{c}=0$, using the parametrization (\ref{monpar}), a straightforward analysis shows that a BPS bound state exists if and only if
\begin{equation} \label{existenseq}
 \tilde{Q}_0 < \tilde{q}_1 < \frac{1}{6}  \qquad {\rm or} \qquad
 \frac{1}{6} < \tilde{q}_1 < \tilde{Q}_0 \, ,
\end{equation}
where $\tilde{q}_1 = \frac{\tilde{k}^2}{2} - b$, and
\begin{equation}
  {\rm sign} \; \tilde{k} = {\rm sign} \; p_0 \, , \qquad \tilde{\cal D} = \frac{8}{9} \tilde{b}^3 - \tilde{n}^2 \geq 0 \, .
\end{equation}
The equilibrium distance (\ref{reqformula}) expressed in rescaled variables is $\tilde{r}_{\rm eq} = \frac{1}{\sqrt{3}} \frac{\tilde{Q}_0-\tilde{q}_1}{\tilde{q}_1 - 1/6}$. The boundary $\tilde{\cal D}=0$ corresponds to vanishing probe entropy, the boundary $\tilde{q}_1 = \tilde{Q}_0$ to a vanishing bound state radius and hence absorption of the probe by the background black hole, and finally the boundary $\tilde{q}_1 = \frac{1}{6}$ corresponds to an infinite bound state radius and hence to decay at marginal stability. This is brought in a more conventional form by returning to the non-scaled variables, which turns the existence condition (\ref{existenseq}) into $\frac{6 Q_0}{P_1} < \frac{6 q_1}{p_0} < y_0^2$ or $\frac{6 Q_0}{P_1} > \frac{6 q_1}{p_0} > y_0^2$. The absorption wall is then clearly seen to correspond to a vanishing probe-background symplectic product $\langle \gamma,\Gamma \rangle = 0$, while the marginal stability wall is at $y_0 = \sqrt{\frac{6 q_1}{p_0}}$.

Notice there exist bound states for all values of $\tilde{Q}_0 \geq 0$ except $1/6$. The number of possible bound states will not be constant however. In particular when $\tilde{Q}_0 \to 1/6$ (or equivalently $y_0 \to y_\star$, where $y_\star= \frac{6 Q_0}{P_1}$ is the attractor fixed point (\ref{attractorpoint}) of the background black hole), the allowed region in the probe charge space shrinks to zero. In appendix \ref{sec:counting} we compute the number of probe bound states, allowing multiple probes with different charges and taking into account the lowest Landau level degeneracies due to the magnetic interaction between the background black hole and the probe charges (but ignoring mutual magnetic interactions between the probes themselves). We do \emph{not} count the internal microstates of the black holes. The logarithm of the number of configurations defined in this way is thus the analog of the notion of configurational entropy in the theory of glasses \cite{TAP,braymoore,Monasson}. The final result for the number ${\cal N}(\epsilon)$ of such configurations with total probe mass over black hole mass less than $\epsilon$ is given by equation (\ref{logNdinal}):
\begin{equation}  \label{logNQCP}
 \log {\cal N}(\epsilon) \sim \epsilon^{5/6} \left|y_0^2 - y_\star^2 \right|^{1/3}  
 \frac{P_1}{y_0^{1/6}} \, ,
\end{equation}
with $y_\star= \frac{6 Q_0}{P_1}$, as in (\ref{attractorpoint}). Thus we see that the number of allowed configurations indeed goes to zero when the critical point $y_0 = y_\star$ is approached, with a nontrivial scaling exponent $1/3$.

It is interesting that even this restricted counting already gives an exponential growth of the number of configurations ${\cal N}$ in the $P_1 \to \infty$ thermodynamic limit. This is typical for glasses \cite{TAP,braymoore,Monasson}. The growth is not as fast as the black hole entropy itself (it is at most $P_1^{3/2}$, if we allow going to the boundary of the probe regime, cf.\ eq.\ (\ref{namfoh})), but the exponentially large number of configurations should nevertheless have important consequences for the thermodynamics of this system.


\subsection{Hot black molecules}

For nonsupersymmetric black holes the analysis becomes more complicated, requiring some numerical assistance to scan the space of possible probe bound states for given $(\tilde{Q}_0,\tilde{c})$. The results of this work are summarized in fig.\ \ref{ST}. We identify four different regions:

{\bf 1}. In the grey region filling the high temperature region, no molecular bound states of any kind exist, as gravity overpowers all other forces.

{\bf 2}. In the yellow regions right below it, bound states exist for some probe charges, but all of them have positive energy at their minimum, so they are metastable (recall we put the zero of the probe potential at the horizon). An example is potential (2) in fig. \ref{potentials}. When approaching the grey-yellow boundary, the minima become higher, are pushed to large radii and become very shallow, while the number of probe configurations goes to zero. This should give a scaling law analogous to (\ref{logNQCP}) but we did not try to extract the scaling exponent. The grey region touches the $T=0$ axis at the quantum critical points $\tilde{Q}_0 = \pm \frac{1}{6} \approx 0.1667$. For $\tilde{Q}_0 > - \frac{1}{6}$, we numerically observed with high accuracy that the probe particles forming the lowest energy bound states are always zero entropy $b=n=0$, $k\neq 0$ particles. In IIA language these are pure D6-branes with $U(1)$ flux, which uplift in M-theory to smooth ``bubbling'' geometries \cite{benawarner,benawarner2,benawarner3,benawarner4,benawarner5,benawarner6,miranda}. In particular the bound state surviving the longest when $c$ is increased is of this type. For $\tilde{Q}_0 < - \frac{1}{6}$ this is no longer the case, and in fact there are no bound states of this type for $-\frac{3}{2} < \tilde{Q}_0 < - \frac{1}{6}$. The blue overlay shows where they reappear in the region $\tilde{Q}_0 < -\frac{3}{2}$. 

{\bf 3}. In the green regions enclosed in the yellow, negative energy bound states exist. Such bound states are energetically stable against tunneling of the probe into the black hole or out to infinity. The green line along the positive $\tilde{Q}_0$ axis represents the BPS bound states discussed in section \ref{sec:susyBS}, which have zero energy.  The negative $\tilde{Q}_0$ axis represents extremal nonsupersymmetric bound states. There is no BPS bound forbidding negative energy states, and by the rule that everything that is not forbidden is allowed, we find indeed that a large subset has deep negative energy minima. 
Interestingly, due to the transient dipping effect described at the end of \ref{sec:bs}, there is a small but finite region on the $\tilde{Q}_0 > 0$ (i.e.\ BPS) side at finite temperature where negative energy bound states exist. It extends to $\tilde{Q}_0 = \frac{1}{54} \approx 0.0185$ (see end of next paragraph).

{\bf 4}. The orange overlay is the region where the background black hole itself is unstable to emission of particles to spatial infinity. We take such emissions to be possible whenever there exists some probe charge such that the probe potential becomes negative at spatial infinity. Notice that the red region includes the green region. Hence whenever a bound state exists that is stable against tunneling of the probe out of its minimum, the background black hole will be unstable to emission of particles. We note, however, that whenever the probe potential exhibits a minimum, it is always found to be lower than the value of the potential at infinity. The destabilizing probe type kicking in first (at the highest $c$) is again a pure fluxed D6 brane, $b=n=0$, over the entire range of $\tilde{Q}_0$ we scanned. This allows computing the boundary of the red region analytically as the value of $c$ for which the asymptotic value of $V_p$ has a double zero viewed as a function of $k$:
\begin{equation}
 c_{\rm BH \, stab} = \frac{1}{4 \sqrt{3} } \frac{3 \Delta -1}{\sqrt{(\Delta +1) (\Delta +5)}} \, , \qquad \Delta \equiv \sqrt{1- 48 \, \tilde{Q}_0} \, .
\end{equation}
The critical line reaches zero at $\Delta=\frac{1}{3}$, i.e.\ $\tilde{Q}_0 = \frac{1}{54}$, which is numerically seen to coincide with the edge of the green region. It asymptotes for $\tilde{Q}_0 \to -\infty$ to $\frac{\sqrt{3}}{4} \approx 0.433$.

\subsection{Bound states in a box}

\begin{figure}
\begin{center}
\includegraphics[height=7.7cm]{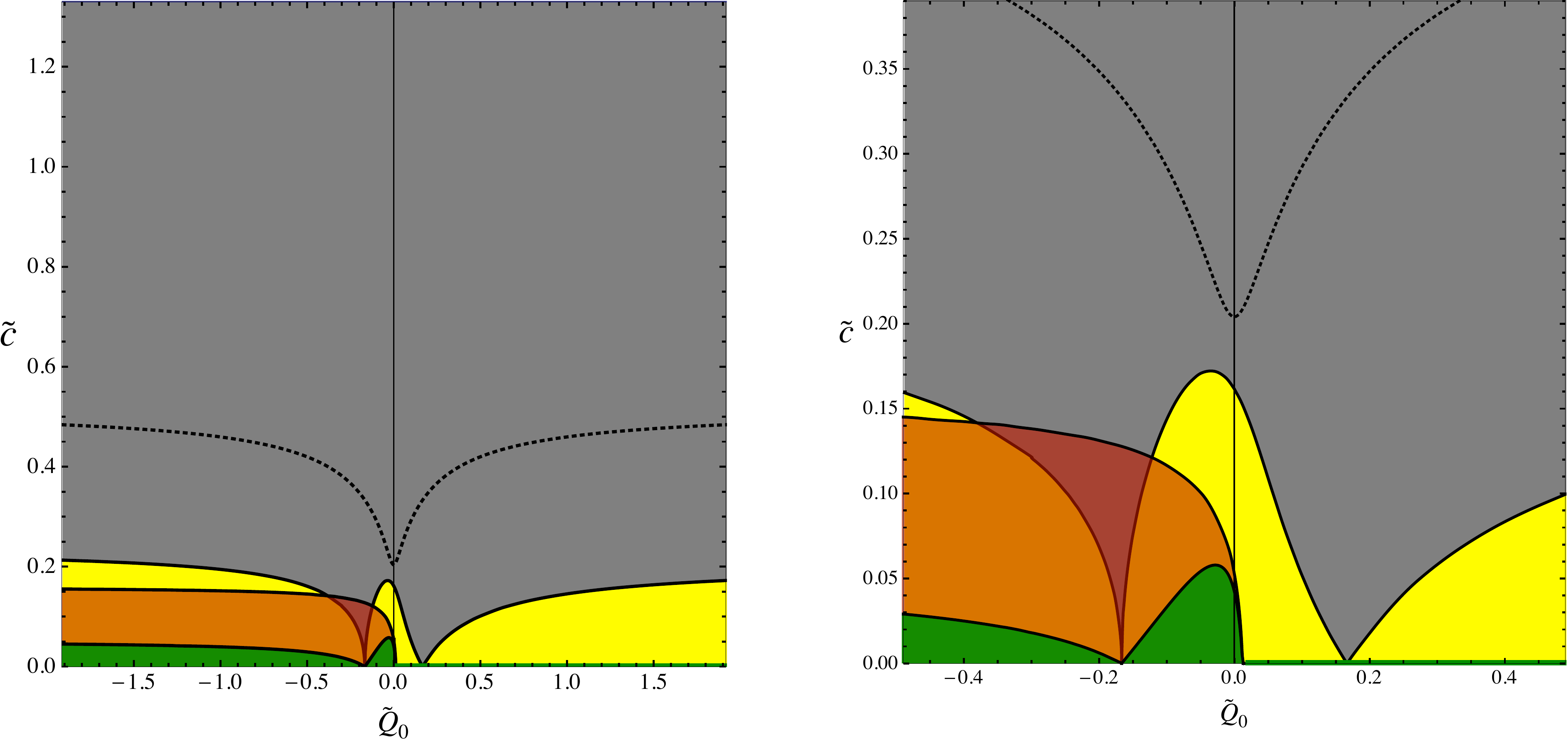}
\end{center}
\vskip-5mm \caption{Same as fig.\ \ref{ST}, but for the system confined in a box, which can be viewed as a rough model for AdS. The box is implemented as a cutoff at radius $\frac{\tilde{c}}{\sinh(\tilde{c} \tilde{\tau})}=1$, i.e.\ $u=1/2$ (with $u$ defined in fig. \ref{potentials}). Bound states with equilibrium positions at $u>1/2$ are discarded, and decay is defined as emission to $u=1/2$. 
 \label{STbox}}
\end{figure}

The black holes under consideration live in asymptotically flat space. Non-BPS black holes are unstable due to Hawking radiation and possibly other emission instabilities, and hot flat space is unstable due to nucleation of black holes \cite{Hawking:1976de,Gross:1982cv}. As a result, it is hard to make sense of this infinitely large system as a statistical mechanical model. To make it better defined, we can put the system in a finite box, either by imposing a cutoff by hand at some finite radius or by embedding the system in AdS$_4$, along the lines of \cite{Page:1981an,Hawking:1982dh,Chamblin:1999tk,Carlip:2003ne}, allowing the black hole to achieve thermal equilibrium with its environment in the box. In simple setups that do not give rise to black molecules, one sees that depending on the size of the box and the temperature, the final equilibrium state can either be a big black hole , or an ordinary thermal gas.\footnote{In the context of AdS$_4$ this transition is nothing more than the Hawking-Page transition \cite{Hawking:1982dh}.} In the case at hand, we may expect much more complicated, glass-like behavior at sufficiently low temperatures, due to the exponentially large number of complex stationary configurations that exist.

Implementing the constraint of putting the system in a box by hand is not too hard. For example we could cut off space at some fixed radial coordinate distance, say $\tilde{r} \equiv \frac{\tilde{c}}{\sinh(\tilde{c} \tilde{\tau})} = 1$ (that is $r=\sqrt{y_0} P_1$ in unrescaled variables), and hold the system at a fixed temperature, fixed $y_0$, and fixed total charge. We can then repeat the existence and stability analysis done for the asymptotically flat case. This is shown in fig.\ \ref{STbox}, the boxed analog of fig.\ \ref{ST} (not plotting the blue region). All the qualitative features remain intact, except that the phase boundaries are pushed down significantly. In fact, they are pushed down below the dotted line indicating the maximal temperature $\tilde{T}$ as a function of $\tilde{c}$ at fixed $\tilde{Q}_0$. Thus, all bound states occur in the region of positive specific heat, and we can say that at sufficiently high temperature, there is only the black hole solution, and it is stable. For the infinite system we could only make the analogous statement for sufficiently high nonextremality parameter (mass) of the black hole. If we imagine the existence of some holographic dual field theory description of the system under consideration, with the black hole representing the unique disordered high temperature thermodynamic equilibrium state and the multiple bound state configurations as the (meta)stable thermodynamic states characteristic for glassy systems below their critical temperature, then a positive specific heat throughout the parameter regime of interest is certainly an expected feature.

\subsection{Thermodynamics and phase structure}

Figures \ref{ST} and \ref{STbox} are certainly suggestive of an interesting phase structure, with the attractor points of the extremal black holes at $(\tilde{Q}_0,\tilde{c})=(\pm \frac{1}{6},0)$ corresponding to quantum critical points, the grey regions corresponding to the ``normal'' disordered high temperature phase and the colored regions to glass-like phases.

However, to say anything definitive about phase structure of this system, a more careful study of the thermodynamic weight of various configurations is needed, as well as an analysis of candidate order parameters and how they scale near phase boundaries and near the critical points. We will leave a full analysis to future work and restrict ourselves here to some simple observations.

At fixed temperature, a thermodynamic system will try to minimize its free energy $F=E-TS$. The free energy satisfies the first law
\begin{equation}
 \delta F = - S \delta T + \delta W \, ,
\end{equation} 
where $\delta W$ denotes the work delivered to the system in some infinitesimal process during which the temperature changes by an infinitesimal amount $\delta T$. In particular this means that if we move a probe particle from some position $r=R$ into the black hole, while keeping the temperature fixed, the total change in free energy of the system $\delta F = \delta F_{\rm BH} + \delta F_{\rm p}$ must be equal to $\delta W = V_{\rm p}(\rm horizon) - V_p(R)$, with $V_{\rm p}=V_{\rm g}+V_{\rm em}$ the probe potential defined in (\ref{probepotdef}).\footnote{Recall we are treating the probe  as a structureless object without internal thermal energy. See footnote \ref{nothermal} for the argument of why this is justified in the probe limit even if the probe itself is a black hole. The argument extends to thermal kinetic energy of the probe moving around in its potential well (provided this lies at finite rescaled radius), which we thus ignore as well.} Since $V_{\rm p}(\rm horizon) = 0$, this means
\begin{equation}
 F(\mbox{probe at R}) - F(\mbox{probe in BH})  = V_{\rm p}(r=R) \, ,
\end{equation}
and thus at fixed temperature the configuration with the probe outside of the black hole will be thermodynamically favored over the one with the probe inside simply whenever $V_{\rm p}(R)$ is negative.

As a check, we note that for $R=\infty$, this is the statement that $\delta F_{\rm BH} + \delta F_{\rm p}=-V_{\rm g}(\infty)-V_{\rm em}(\infty)$. Since $\delta F_{\rm p} = - m_{\rm p} = - V_{\rm g}(\infty)$ (as the probe is first at infinity and then gone, and we are allowed to ignore as before thermal contributions to the probe free energy), this is equivalent to $\delta F_{\rm BH} = - V_{\rm em}$, which can be directly checked from the expressions in \ref{sec:MET}, without using the first law, taking care to vary $c$ at the same time as the charges to keep $T$ fixed. 

To conclude, figures \ref{ST} and \ref{STbox} give information about the thermodynamic preferences of the system for perturbations around the single black hole state. For example when in the green region, the black hole will start to populate black hole halos of many different charge types. We should keep in mind however that as soon as the number of such probes becomes macroscopic,  or when they coalesce into black holes of size comparable to the background black hole, our neglect of the thermal internal, kinetic and inter-probe interaction energies is no longer justified. Hence we cannot read off the endpoint of this evolution from the diagrams. Nevertheless, the existence of exponentially many bound states with free energy below the black hole free energy strongly suggests glass-like behavior. This is not entirely obvious though, since the plots also show that the black hole core of any negative energy bound state is always unstable to emission of particles to infinity or to the boundary of the confining box, so for a sufficiently large box, it is conceivable that the final equilibrium state may still be a simple dilute charged gas. We will  revisit such questions in future work.

Finally, let us comment on the tentative interpretation of the points $\tilde{Q}_0 = \pm \frac{1}{6}$ at $T=0$ as quantum critical points. As we mentioned several times already, they correspond to the case in which the scalars at infinity are at the black hole attractor point, $y_0 = y_\star= \sqrt{\frac{6 |Q_0|}{|P_1|}}$. In this case, the background scalars do not flow but rather are constant over all of space, and no molecular bound state configurations exist -- everything has been sucked into the black hole or pushed away to infinity. (It is known that this remains true away from the probe limit, at least in the BPS case: non-marginal bound states cannot exist at the attractor point of the total charge.) At zero temperature, the geometry is that of an extremal Reissner-Nordstr\"om black hole, developing in particular an AdS$_2 \times S^2$ throat, suggestive of a holographic conformal fixed point. 

Taking $y_0$ away from the attractor point but still keeping $T=0$, the AdS$_2$ is preserved but the scalars will now flow from $y_0$ at infinity to $y_\star$ at the horizon. Putting the system in a box of finite radius $R$ and decreasing $R$ while keeping $y_0$ fixed will have roughly the same effect as moving $y_0$ along the attractor flow towards the fixed point $y_\star$; in particular near $\tilde{Q}_0 = \frac{1}{6}$, the configurational entropy will decrease by some power of $R$ dictated by (\ref{logNQCP}). In a hypothetical field theory dual, regardless of the large $r$ asymptotics of spacetime \cite{Bredberg:2010ky}, decreasing $R$ would correspond to flowing to the IR, so this would correspond to a power law decrease of the configurational entropy when coarse graining over increasing length scales. 


If $|\tilde{Q}_0| = \frac{1}{6}$ is to be a quantum critical point associated to a quantum phase transition, then we should see different physical properties on the two sides of it. Near the supersymmetric critical point, one such property can be inferred by inspecting (\ref{existenseq}). Recall that pulling a probe out of the background black hole produces a D6 magnetic dipole moment pointing along the radial direction, equal to $\mu=p_0 r_{\rm eq} = p_0 \sqrt{y_0} P_1  \frac{1}{\sqrt{3}} \frac{\tilde{Q}_0-\tilde{q}_1}{\tilde{q}_1 - \frac{1}{6}}$, as well as an angular momentum $J = \frac{1}{2} \langle \gamma,\Gamma \rangle = \frac{1}{2} P_1 y_0^2 p_0 (\tilde{Q}_0 - \tilde{q}_1)$. Hence
\begin{equation} \label{gequation}
 \vec{J} = g \vec{\mu} \, ,  \qquad g \equiv \tfrac{\sqrt{3}}{{2}}   y_0^{3/2} \left(\tilde{q}_1 - \tfrac{1}{6}\right) \, . 
\end{equation}     
From (\ref{existenseq}) we see that the range of possible values of $|g|$ runs from 0 to a maximal value proportional to $|\tilde{Q}_0 - \frac{1}{6}|$, and that it changes sign across the phase boundaries. The coefficient $g$ is a physical observable, and could be used as an order parameter to distinguish the two putative phases.



\vskip10mm

\noindent {\bf \Large Acknowledgements} 

\vskip5mm

\noindent We would like to thank Iosif Bena, Hyeyoun Chung, Clay Cordova, Michael Douglas, Hajar Ebrahim, Sean Hartnoll, Greg Moore, Pavel Petrov, Andy Strominger, Dieter Van den Bleeken and Bert Vercnocke for valuable input and discussions. We also thank the Aspen Center for Physics and the island of Hydra for providing the perfect environment for completing this paper. This work was supported in part by DOE grant DE-FG02-91ER40654, NSF grant no. 0756174 and by a grant of the John Templeton Foundation. The opinions expressed in this publication are those of the authors and do not necessarily reflect the views of the John Templeton Foundation.

\appendix

\section{Counting configurations} \label{sec:counting}

There exist supersymmetric probe bound state solutions for all values of $\tilde{Q}_0 \geq 0$ except $1/6$. The number of possible bound state solutions will not be constant however. In particular when $\tilde{Q}_0 \to 1/6$, the allowed region in the probe charge space shrinks to zero. In this appendix we obtain an estimate for the number of supersymmetric probe bound state solutions near this point. First we consider the case in which we simply count the number of allowed probe charges, later on we will include the lowest Landau level degeneracies for each choice of probe.

\subsection{Single probe}

For a single probe, this number is ${\cal N}_1 = \sum_{\gamma \in A} 1$, where $A$ is the allowed region in probe charge space, bounded by the requirement that the bound state exists and the probe approximation is satisfied. As we take the limit $P_1 \to \infty$, $A$ will contain an increasingly large number of charges, hence the number of lattice points contained in $A$ can be estimated by computing the volume of $A$ in charge space:
\begin{equation} \label{countintegral}
 {\cal N}_1 \approx \int_A dp_0 \, dp_1 \, dq_1 \, dq_0 \, 1 \,
 = \int dp_0 \, p_0^3 y_0^6 \int_{\tilde{A}} d \tilde{b} \, d \tilde{n} \, d\tilde{k} \, . 
\end{equation}
Here $\tilde{A}$ is the region of allowed values of $(\tilde{k},\tilde{n},\tilde{b})$. Since $|\tilde{k}|$ is bounded by $\sqrt{2(\tilde{Q}_0+\tilde{b})}$ and $\sqrt{2(\frac{1}{6}+\tilde{b})}$ according to  (\ref{existenseq}), the integral over $\tilde{k}$ gives a factor $f_1(\tilde{b}) \sim \left| \sqrt{\frac{1}{6} + \tilde{b}} - \sqrt{\tilde{Q}_0 + \tilde{b}} \right|$. (We drop  irrelevant numerical factors here and in what follows.) Because of the constraint $\tilde{n}^2 \leq \frac{8}{9} \tilde{b}^3$, the integral over $\tilde{n}$ gives a factor $f_2(\tilde{b}) \sim \tilde{b}^{3/2}$. Performing the integral over $\tilde{b}$ gives
\begin{equation}
 \int d\tilde{b} \, f_1(\tilde{b}) \, f_2(\tilde{b}) \sim |\tilde{Q}_0 - \tfrac{1}{6}| \,\tilde{b}_{\rm max}^2 \, .
\end{equation}
Here $\tilde{b}_{\rm max}^2$ is the maximal value of $\tilde{b}$, which we have assumed to be very large so we are allowed to drop terms of order $\tilde{b}_{\rm max}$ and lower. The reason $b_{\rm max}$ is not infinite is the requirement that the probe approximation should be valid. To estimate it, recall that $m_p/M = \frac{y_0 p_0}{P_1} \frac{\tilde{m}_p}{\tilde{M}}$. Since we are exploring the region $\tilde{Q}_0 \approx 1/6$, $\tilde{M}$ is of order 1. For large $\tilde{b}$, $\tilde{m}_p \sim \tilde{q}_0 \sim \tilde{b}^{3/2}$, so the probe approximation requires $\tilde{b}_{\rm max} = \left(\epsilon \frac{P_1}{y_0 p_0}\right)^\alpha$, where $\alpha=2/3$ and $\epsilon \sim m_p/M$ is some small number, the maximal $m_p/M$ we allow. We are then left with the integral over $p_0$:
\begin{eqnarray} \label{N1result}
 {\cal N}_1(\tfrac{m_p}{M} < \epsilon) &\sim& \epsilon^{2\alpha} \left|\tilde{Q}_0 - \tfrac{1}{6}\right| \int_0^{\epsilon P_1/y_0} dp_0 \, p_0^{3-2\alpha} y_0^{6-2\alpha} P_1^{2\alpha} \, \\
 &\sim&  \epsilon^{4} \left|\tilde{Q}_0 - \tfrac{1}{6}\right| y_0^2 P_1^4 \, \\
 &=& \epsilon^4 \left|\tfrac{6 Q_0}{P_1} - y_0^2 \right| P_1^4 \, . \label{CN1}
\end{eqnarray}
The upper integral bound $p_0 < \epsilon P_1/y_0$ comes again from requiring that we remain within the probe approximation, this time in the limit of large $p_0$. Notice that the final result does not depend on the actual value of $\alpha$. In fact, we could have inferred the prefactor simply from the scaling symmetries of the system: From (\ref{countintegral}) it is clear that ${\cal N}_1$ has scaling weights $(0,6,4)$ under the symmetries of \ref{sec:allscalings}, while $\epsilon$, $P_1$ and $y_0$ have scaling weights $(-1,0,1)$, $(1,1,0)$ and $(0,1,0)$, respectively. This uniquely determines their powers.


The above expressions are valid when $\left|\tilde{Q}_0 - \tfrac{1}{6}\right| \ll 1$, i.e.\ $\left|\frac{6 Q_0}{P_1} - y_0^2 \right| \ll \frac{Q_0}{P_1}$, or in other words close to the attractor point: $y_0 \to y_\star = \sqrt{\frac{6 Q_0}{P_1}}$. Notice that since the probe approximation requires $y_0 \ll P_1$ (as discussed in section \ref{sec:allscalings}), self-consistency in this regime requires $Q_0 \ll P_1^3$, i.e.\ we are necessarily in the non-Cardy regime.

Pushing $y_0$ and $\epsilon$ as high up as possible while conceivably still yielding more or less sensible results, i.e.\ 
\begin{equation} \label{namfoh}
  \qquad y_0 \sim P_1 \, , \qquad  Q_0 \sim P_1^3\, , \qquad \epsilon \sim 1 \, , 
\end{equation} 
we get
\begin{equation} \label{maxN1}
 {\cal N}_{1,\rm max} \sim P_1^6 \, .
\end{equation}

\subsection{Multiple probes}

In the probe approximation we can also easily build multi-probe bound states: By assumption, the probes do not backreact so we can simply superimpose the single probe configurations, as long as we keep the total probe mass $\sum_i m_{p i}$ small compared to $M$.\footnote{In general the probes will interact with each other, with interaction strength given by their mutual symplectic products. For probes which happen to have small symplectic products with the background black hole, i.e.\ probes which are close to be swallowed by the black hole, these interactions become important even in the probe limit. We will ignore such boundary cases here.} Imagine a general situation in which the number of single probe states with $m_p/M < \epsilon$ is given by
\begin{equation}
 {\cal N}_1(\epsilon) = A \epsilon^n \, , 
\end{equation}
where $A$ is some large number. In the case at hand $n=4$, but we will keep things slightly more general here for future reference. The density of single particle states at $\frac{m_p}{M} = \epsilon$ is then $d{\cal N}_1(\epsilon) = n \epsilon^{n-1} A$. For $K$ labeled probes, the density of states at $\frac{m_i}{M} = \epsilon_i$, $i=1,\ldots,K$ is 
\begin{equation}
 d {\cal N}_K (\epsilon_1,\ldots,\epsilon_K) = (n A)^K \epsilon_1^{n-1} \cdots \epsilon_K^{n-1} \, d\epsilon_1 \cdots d \epsilon_K \, .
\end{equation}
The total number of states with an arbitrary number $K$ of unlabeled probes satisfying $\sum_i \frac{m_i}{M} < \epsilon$ is therefore
\begin{equation}
 {\cal N}(\epsilon) = \sum_{K=0}^\infty \frac{(n A)^K}{K!} \int_{\sum_i \epsilon_i < \epsilon} d\epsilon_1 \cdots d\epsilon_K \, \epsilon_1^{n-1} \cdots \epsilon_K^{n-1} \, .
\end{equation}
The $1/K!$ corrects in a classical way for overcounting.\footnote{We do not use quantum statistics because the probability that two probes occupy the same quantum microstate is completely negligible in this setup. This is already true if the probes are considered to be point particles without internal degrees of freedom, but becomes obvious without work when one takes into account the huge number of internal microstates the individual probe black holes can choose from.} 
The integral can be factorized by representing the constraint $\sum_i \epsilon_i < \epsilon$ as the contour integral $\frac{1}{2 \pi i} \int \frac{d\lambda}{\lambda} e^{\lambda(\epsilon-\sum_i \epsilon_i)}$, where the contour is taken to be on the right of the pole at $\lambda=0$. This yields
\begin{eqnarray}
 {\cal N}(\epsilon) &=& \frac{1}{2 \pi i} \int \frac{d\lambda}{\lambda} \, e^{\lambda \epsilon}
  \sum_{K=0}^\infty \frac{(n A)^K}{K!} \left( \int d\epsilon_1 \, \epsilon_1^{n-1} \, e^{-\lambda \epsilon_1} \right)^K \nonumber \\
  &=& \frac{1}{2 \pi i} \int \frac{d\lambda}{\lambda} \, \exp \left( \lambda \epsilon + \frac{n! A}{\lambda^{n}} \right) \, .
\end{eqnarray}
At large $A$ this can be computed by saddle point evaluation. To leading order, dropping order one numerical factors: 
\begin{equation} \label{multiprobefin}
 \log {\cal N}(\epsilon) \sim \left( A \epsilon^n \right)^{\frac{1}{n+1}} \, .
\end{equation}
Applying this to (\ref{CN1}) gives for the total number of distinct probe configurations 
\begin{equation}
 {\cal N}(\epsilon) \approx \exp \left( \kappa \, \epsilon^{4/5} \left|y_\star^2 - y_0^2 \right|^{1/5} P_1^{4/5} \right) \, ,
\end{equation}
where $\kappa$ is some order 1 constant and $y_\star = \tfrac{6 Q_0}{P_1}$ is the attractor fixed point.

In the regime (\ref{namfoh}), we thus get a configurational entropy
\begin{equation}
 \log {\cal N}_{\rm max} \sim P_1^{6/5} \, .
\end{equation}

\subsection{Including Landau level degeneracies}

A single probe bound to the background black hole has classically an $S^2$ moduli space, but because a magnetic field threads the sphere, the space of quantum BPS ground states (i.e.\ the lowest Landau level) will be degenerate. The degeneracy is given by the effective magnetic flux as seen by the probe \cite{Denef:2002ru}:
\begin{equation}
 d_\gamma = |p_0 Q_0 - q_1 P_1| = p_0 P_1 y_0^2 | \tilde{Q}_0 - \tilde{q}_1 | \, .
\end{equation}
To count the total number of such one particle ground states (ignoring the internal degrees of freedom of the probe and of the black hole), it suffices to replace the 1 in (\ref{countintegral}) by the LLL degeneracy factor $d_\gamma$. Because this has scaling dimensions $(1,3,1)$, doing so will add an additional factor $\epsilon y_0 P_1^2$ to the final result (\ref{CN1}). Furthermore, because the insertion $|\tilde{Q}_0 - \tilde{q}_1|$ is of generically of order $|\tilde{Q}_0 - \frac{1}{6}|$ over the integration domain, its effect will be to modify the power of $|\tilde{Q}_0 - \frac{1}{6}|$ from linear to quadratic. All in all we get
\begin{eqnarray}
 {\cal N}_{{\rm LLL},1}(\epsilon) &\sim&  \epsilon^{5} \left|\tilde{Q}_0 - \tfrac{1}{6}\right|^2 y_0^3 P_1^6 \, \\
 &=& \epsilon^5 \left|\tfrac{6 Q_0}{P_1} - y_0^2 \right|^2 \frac{P_1^6}{y_0} \, . \label{CNf} 
\end{eqnarray}
For the multi-probe system we get from this, using (\ref{multiprobefin}),
\begin{equation} \label{logNdinal}
 \log {\cal N}_{\rm LLL}(\epsilon) \sim \epsilon^{5/6} \left|\tfrac{6 Q_0}{P_1} - y_0^2 \right|^{1/3}  
 \frac{P_1}{y_0^{1/6}} 
\end{equation}
Finally, in regime (\ref{namfoh}), we get a configurational entropy
\begin{equation}
 \log {\cal N}_{\rm LLL, \, max} \sim P_1^{3/2} \, .
\end{equation}
Recall that the entropy of the black hole scales as $P_1^3$.


\begin{thebibliography}{93}

 \bibitem{Weyl}
  H.~Weyl,
  ``The theory of gravitation,''
  {\it Ann. Physik}, {\bf 54}, 117 (1917)

\bibitem{majum}
  S.~D.~Majumdar,
  ``A class of exact solutions of Einstein's field equations,''
  Phys.\ Rev.\  {\bf 72}, 390-398 (1947).
  

\bibitem{pap}   A.~Papapetrou,
  ``A Static solution of the equations of the gravitational field for an arbitrary charge distribution,''
  Proc.\ Roy.\ Irish Acad.\ (Sect.\ A) {\bf A51}, 191-204 (1947).

  \bibitem{Ehlers}
  J.~Ehlers and W.~Kundt,
  {\it Gravitation: an introduction to current research},
  ed. L. Witten. Wiley, New York (1962)

  \bibitem{Israel}
  W.~Israel and K.~A.~Khan,
  ``Collinear particles and bondi dipoles in general relativity,"
  {\it Il Nuovo Cimento}, Vol. 33, Number {\bf{2}}, 331-344 (1964)

  \bibitem{Kinnersley}
    W.~Kinnersley, M.~Walker,
  ``Uniformly accelerating charged mass in general relativity,''
  Phys.\ Rev.\  {\bf D2}, 1359-1370 (1970).
  

  \bibitem{Bonnor}
  W.~B.~Bonnor,
  ``The sources of the vacuumC-metric,''
  {\it Gen. Rel. Grav.} {\bf 15}, 535-551 (1983)


\bibitem{Kastor:1992nn}
  D.~Kastor, J.~H.~Traschen,
  ``Cosmological multi - black hole solutions,''
  Phys.\ Rev.\  {\bf D47}, 5370-5375 (1993).
  [hep-th/9212035].

\bibitem{SabraLust}
  K.~Behrndt, D.~Lust, W.~A.~Sabra,
  ``Stationary solutions of N=2 supergravity,''
  Nucl.\ Phys.\  {\bf B510}, 264-288 (1998).
  [hep-th/9705169].

\bibitem{Booth:1998gf}
  I.~S.~Booth, R.~B.~Mann,
  ``Cosmological pair production of charged and rotating black holes,''
  Nucl.\ Phys.\  {\bf B539}, 267-306 (1999).
  [gr-qc/9806056].


\bibitem{Denef:2000nb}
  F.~Denef,
  ``Supergravity flows and D-brane stability,''
  JHEP {\bf 0008}, 050 (2000).
  [hep-th/0005049].  
  
\bibitem{LopesCardoso:2000qm}
  G.~Lopes Cardoso, B.~de Wit, J.~Kappeli, T.~Mohaupt,
  ``Stationary BPS solutions in N=2 supergravity with R**2 interactions,''
  JHEP {\bf 0012}, 019 (2000).
  [hep-th/0009234].
  
  
\bibitem{Denef:2002ru}
  F.~Denef,
  ``Quantum quivers and Hall / hole halos,''
  JHEP {\bf 0210}, 023 (2002).
  [hep-th/0206072].
  
\bibitem{Bates:2003vx}
  B.~Bates, F.~Denef,
  ``Exact solutions for supersymmetric stationary black hole composites,''
  [hep-th/0304094].

\bibitem{Tan:2003jz}
  H.~S.~Tan and E.~Teo,
  ``Multi - black hole solutions in five-dimensions,''
  Phys.\ Rev.\  D {\bf 68}, 044021 (2003)
  [arXiv:hep-th/0306044].

\bibitem{Jejjala:2005yu}
  V.~Jejjala, O.~Madden, S.~F.~Ross, G.~Titchener,
  ``Non-supersymmetric smooth geometries and D1-D5-P bound states,''
  Phys.\ Rev.\  {\bf D71}, 124030 (2005).
  [hep-th/0504181].

\bibitem{Chng:2006gh}
  B.~Chng, R.~B.~Mann, C.~Stelea,
  ``Accelerating Taub-NUT and Eguchi-Hanson solitons in four dimensions,''
  Phys.\ Rev.\  {\bf D74}, 084031 (2006).
  [gr-qc/0608092].

\bibitem{Griffiths:2006tk}
  J.~B.~Griffiths, P.~Krtous and J.~Podolsky,
  ``Interpreting the C-metric,''
  Class.\ Quant.\ Grav.\  {\bf 23}, 6745 (2006)
  [arXiv:gr-qc/0609056].

\bibitem{Ishihara:2006iv}
  H.~Ishihara, M.~Kimura, K.~Matsuno and S.~Tomizawa,
  ``Kaluza-Klein Multi-Black Holes in Five-Dimensional Einstein-Maxwell
  Theory,''
  Class.\ Quant.\ Grav.\  {\bf 23}, 6919 (2006)
  [arXiv:hep-th/0605030].


\bibitem{Giusto:2007tt}
  S.~Giusto, S.~F.~Ross and A.~Saxena,
  ``Non-supersymmetric microstates of the D1-D5-KK system,''
  JHEP {\bf 0712}, 065 (2007)
  [arXiv:0708.3845 [hep-th]].

\bibitem{Giusto:2007fx}
  S.~Giusto and A.~Saxena,
  ``Stationary axisymmetric solutions of five dimensional gravity,''
  Class.\ Quant.\ Grav.\  {\bf 24}, 4269 (2007)
  [arXiv:0705.4484 [hep-th]].

\bibitem{Ford:2007th}
  J.~Ford, S.~Giusto, A.~Peet and A.~Saxena,
  ``Reduction without reduction: Adding KK-monopoles to five dimensional
  stationary axisymmetric solutions,''
  Class.\ Quant.\ Grav.\  {\bf 25}, 075014 (2008)
  [arXiv:0708.3823 [hep-th]].



\bibitem{Elvang:2007rd}
  H.~Elvang, P.~Figueras,
  ``Black Saturn,''
  JHEP {\bf 0705}, 050 (2007).
  [hep-th/0701035].

\bibitem{Rogatko:2007kq}
  M.~Rogatko,
  ``First Law of Black Saturn Thermodynamics,''
  Phys.\ Rev.\  D {\bf 75}, 124015 (2007)
  [arXiv:0705.3697 [hep-th]].

\bibitem{Gaiotto:2007ag}
  D.~Gaiotto, W.~Li and M.~Padi,
  ``Non-Supersymmetric Attractor Flow in Symmetric Spaces,''
  JHEP {\bf 0712}, 093 (2007)
  [arXiv:0710.1638 [hep-th]].


\bibitem{Camps:2008hb}
  J.~Camps, R.~Emparan, P.~Figueras, S.~Giusto and A.~Saxena,
  ``Black Rings in Taub-NUT and D0-D6 interactions,''
  JHEP {\bf 0902}, 021 (2009)
  [arXiv:0811.2088 [hep-th]].

\bibitem{Evslin:2008py}
  J.~Evslin and C.~Krishnan,
  ``Metastable Black Saturns,''
  JHEP {\bf 0809}, 003 (2008)
  [arXiv:0804.4575 [hep-th]].

\bibitem{Chng:2008sr}
  B.~Chng, R.~B.~Mann, E.~Radu and C.~Stelea,
  ``Charging Black Saturn?,''
  JHEP {\bf 0812}, 009 (2008)
  [arXiv:0809.0154 [hep-th]].



\bibitem{Bena:2009en}
  I.~Bena, S.~Giusto, C.~Ruef, N.~P.~Warner,
  ``Multi-Center non-BPS Black Holes: the Solution,''
  JHEP {\bf 0911}, 032 (2009).
  [arXiv:0908.2121 [hep-th]].

\bibitem{AlAlawi:2009qe}
  J.~H.~Al-Alawi and S.~F.~Ross,
  ``Spectral Flow of the Non-Supersymmetric Microstates of the D1-D5-KK
  System,''
  JHEP {\bf 0910}, 082 (2009)
  [arXiv:0908.0417 [hep-th]].

\bibitem{Stelea:2009ur}
  C.~Stelea, K.~Schleich and D.~Witt,
  ``Charged Kaluza-Klein double-black holes in five dimensions,''
  Phys.\ Rev.\  D {\bf 83}, 084037 (2011)
  [arXiv:0909.3835 [hep-th]].

\bibitem{Ferrara:2010cw}
  S.~Ferrara, A.~Marrani, E.~Orazi,
  ``Split Attractor Flow in N=2 Minimally Coupled Supergravity,''
  Nucl.\ Phys.\  {\bf B846}, 512-541 (2011).
  [arXiv:1010.2280 [hep-th]].

\bibitem{Emparan:2010sx}
  R.~Emparan, P.~Figueras,
  ``Multi-black rings and the phase diagram of higher-dimensional black holes,''
  JHEP {\bf 1011}, 022 (2010).
  [arXiv:1008.3243 [hep-th]].

\bibitem{Anninos:2010gh}
  D.~Anninos and T.~Anous,
  ``A de Sitter Hoedown,''
  JHEP {\bf 1008}, 131 (2010)
  [arXiv:1002.1717 [hep-th]].


\bibitem{Stelea:2011jm}
  C.~Stelea, C.~Dariescu and M.~A.~Dariescu,
  ``Static charged double-black rings in five dimensions,''
  Phys.\ Rev.\  D {\bf 84}, 044009 (2011)
  [arXiv:1107.3484 [gr-qc]].

\bibitem{Bena:2011zw}
  I.~Bena, B.~D.~Chowdhury, J.~de Boer, S.~El-Showk, M.~Shigemori,
  ``Moulting Black Holes,''
  
  [arXiv:1108.0411 [hep-th]].



\bibitem{SteleaNew}
  C.~Stelea, K.~Schleich and D.~Witt,
  ``Non-extremal multi-Kaluza-Klein black holes in five dimensions,"
  [arXiv:1108.5145 [hep-th]].

\bibitem{Andriyash:2010qv}
  E.~Andriyash, F.~Denef, D.~L.~Jafferis, G.~W.~Moore,
  ``Wall-crossing from supersymmetric galaxies,''
  [arXiv:1008.0030 [hep-th]].

\bibitem{Adams:2011rj}
  A.~Adams, S.~Yaida,
  ``Disordered Holographic Systems I: Functional Renormalization,''  
  [arXiv:1102.2892 [hep-th]].

\bibitem{Kachru:2009xf}
  S.~Kachru, A.~Karch, S.~Yaida,
  ``Holographic Lattices, Dimers, and Glasses,''
  Phys.\ Rev.\  {\bf D81}, 026007 (2010).
  [arXiv:0909.2639 [hep-th]].


\bibitem{kurchan}
  J.~Jurchan,
  ``Glasses"~,
  \emph{Seminaire Poincare XIII} (2009)

\bibitem{spinglassbook}
  M.~Mezard, G.~Parisi, and M.~A.~Virasoro, \emph{Spin Glass Theory and Beyond.} vol. 9,
  \emph{Lecture Notes in Physics}, (World Scientic, 1987). ISBN 9971501155. URL http:
  //books.google.com/books?id=ZIF9QgAACAAJ.

\bibitem{amirreview} A.~Amir, Y.~Oreg, Y.~Yospeh, ``Electron Glass Dynamics, ''
	Annual Review of Condensed Matter Physics, vol. 2, p.235-262, arXiv:1010.5767v4, [cond-mat.dis-nn]
     
\bibitem{amirtalk} ``Localization, Anomalous Diffusion and Slow Relaxations in Disordered Systems'', \href{http://online.kitp.ucsb.edu/online/electroglass10/amir/}{http://online.kitp.ucsb.edu/online/electroglass10/amir/}

\bibitem{DuXu} Xu Du, Guohong Li, Eva Y. Andrei, M. Greenblatt, P. Shuk, ``Ageing memory and glassiness of a driven vortex system'', Nature Physics, Volume 3, Issue 2, pp. 111-114 (2007). 
  
\bibitem{icecream} \href{http://en.wikipedia.org/wiki/Ice_cream#Using_liquid_nitrogen}{http://en.wikipedia.org/wiki/Ice cream Using liquid nitrogen} 

\bibitem{glassproblem} P.W. Anderson, ``Through the Glass Lightly'', Science 267:1615 (1995), \\
  D.~Kennedy and C.~Norman, ``What don't we know?'', Science 309:5731 (2005).

\bibitem{Denef:2011ee}
  F.~Denef,
  ``TASI lectures on complex structures,''
  [arXiv:1104.0254 [hep-th]].

  \bibitem{diothesis}
  D.~Anninos,
  ``Classical and Quantum Symmetries of De Sitter Space,"
  Ph.D Thesis, Harvard University (May 2011)

\bibitem{Gomis:2005wc}
  J.~Gomis, F.~Marchesano, D.~Mateos,
  ``An Open string landscape,''
  JHEP {\bf 0511}, 021 (2005).
  [hep-th/0506179].


\bibitem{Denef:2007vg}
  F.~Denef, G.~W.~Moore,
  ``Split states, entropy enigmas, holes and halos,''
  [hep-th/0702146 [HEP-TH]].

\bibitem{Kraus:1994by}
  P.~Kraus, F.~Wilczek,
  ``Selfinteraction correction to black hole radiance,''
  Nucl.\ Phys.\  {\bf B433}, 403-420 (1995).
  [gr-qc/9408003].

\bibitem{Kraus:1994fj}
  P.~Kraus, F.~Wilczek,
  ``Effect of selfinteraction on charged black hole radiance,''
  Nucl.\ Phys.\  {\bf B437}, 231-242 (1995).
  [hep-th/9411219].

\bibitem{KeskiVakkuri:1996xp}
  E.~Keski-Vakkuri and P.~Kraus,
  ``Microcanonical D-branes and back reaction,''
  Nucl.\ Phys.\  B {\bf 491}, 249 (1997)
  [arXiv:hep-th/9610045].

\bibitem{Parikh:1999mf}
  M.~K.~Parikh, F.~Wilczek,
  ``Hawking radiation as tunneling,''
  Phys.\ Rev.\ Lett.\  {\bf 85}, 5042-5045 (2000).
  [arXiv:hep-th/9907001 [hep-th]].

\bibitem{bertthesis}
  B.~Vercnocke,
  ``Hidden Structures of Black Holes,''
  [arXiv:1011.6384 [hep-th]].

\bibitem{jacob} J.~Barandes, ``Special geometry and black holes," to appear.

\bibitem{Galli:2011fq}
  P.~Galli, T.~Ortin, J.~Perz, C.~S.~Shahbazi,
  ``Non-extremal black holes of N=2, d=4 supergravity,''
  JHEP {\bf 1107}, 041 (2011).
  [arXiv:1105.3311 [hep-th]].

\bibitem{deWit}
  B.~de Wit, P.~G.~Lauwers, A.~Van Proeyen,
  ``Lagrangians of N=2 Supergravity - Matter Systems,''
  Nucl.\ Phys.\  {\bf B255}, 569 (1985).

\bibitem{italians}
  L.~Andrianopoli, M.~Bertolini, A.~Ceresole, R.~D'Auria, S.~Ferrara, P.~Fre, T.~Magri,
  ``N=2 supergravity and N=2 superYang-Mills theory on general scalar manifolds: Symplectic covariance, gaugings and the momentum map,''
  J.\ Geom.\ Phys.\  {\bf 23}, 111-189 (1997).
  [arXiv:hep-th/9605032 [hep-th]].

\bibitem{Billo:1999ip}
  M.~Billo, S.~Cacciatori, F.~Denef, P.~Fre, A.~Van Proeyen, D.~Zanon,
  ``The 0-brane action in a general D = 4 supergravity background,''
  Class.\ Quant.\ Grav.\  {\bf 16}, 2335-2358 (1999).
  [hep-th/9902100].

\bibitem{Gibbons:1982ih}
  G.~W.~Gibbons,
  ``Antigravitating Black Hole Solitons with Scalar Hair in N=4 Supergravity,''
  Nucl.\ Phys.\  {\bf B207}, 337-349 (1982).



\bibitem{Ceresole:2007wx}
  A.~Ceresole and G.~Dall'Agata,
  ``Flow Equations for Non-BPS Extremal Black Holes,''
  JHEP {\bf 0703}, 110 (2007)
  [arXiv:hep-th/0702088].

\bibitem{Perz:2008kh}
  J.~Perz, P.~Smyth, T.~Van Riet, B.~Vercnocke,
  ``First-order flow equations for extremal and non-extremal black holes,''
  JHEP {\bf 0903}, 150 (2009).
  [arXiv:0810.1528 [hep-th]].

\bibitem{Ferrara:2008ap}
  S.~Ferrara, A.~Gnecchi and A.~Marrani,
  ``d=4 Attractors, Effective Horizon Radius and Fake Supergravity,''
  Phys.\ Rev.\  D {\bf 78}, 065003 (2008)
  [arXiv:0806.3196 [hep-th]].

\bibitem{Gimon:2009gk}
  E.~G.~Gimon, F.~Larsen, J.~Simon,
  ``Constituent Model of Extremal non-BPS Black Holes,''
  JHEP {\bf 0907}, 052 (2009).
  [arXiv:0903.0719 [hep-th]].


\bibitem{Ferrara:1997tw}
  S.~Ferrara, G.~W.~Gibbons, R.~Kallosh,
  ``Black holes and critical points in moduli space,''
  Nucl.\ Phys.\  {\bf B500}, 75-93 (1997).
  [hep-th/9702103].

\bibitem{Olive:1980sd}
  D.~I.~Olive,
  ``Classical Solutions In Gauge Theories - Spherically Symmetric Monopoles - Lax Pairs And Toda Lattices,''
  In *Bad Honnef 1980, Proceedings, Current Topics In Elementary Particle Physics*, 199-217 and London Imp. Coll. - ICTP-80-81-01 (80,REC.NOV.) 19p.  

\bibitem{Davies:1978mf}
  P.~C.~W.~Davies,
  ``Thermodynamics Of Black Holes,''
  Proc.\ Roy.\ Soc.\ Lond.\  {\bf A353}, 499-521 (1977).

\bibitem{Shmakova:1996nz}
  M.~Shmakova,
  ``Calabi-Yau black holes,''
  Phys.\ Rev.\  {\bf D56}, 540-544 (1997).
  [hep-th/9612076].

\bibitem{deBoer:2008fk}
  J.~de Boer, F.~Denef, S.~El-Showk, I.~Messamah, D.~Van den Bleeken,
  ``Black hole bound states in AdS(3) x S**2,''
  JHEP {\bf 0811}, 050 (2008).
  [arXiv:0802.2257 [hep-th]].

\bibitem{TAP}
   D.~J.~Thouless, P.~W.~Anderson, R.~G.~Palmer,
   {\it{Phil. Mag.}} {\bf{35}} 593 (1977)

\bibitem{braymoore}
   A.~J.~Bray, M.~A.~Moore, 
  ``Metastable States in Spin Glasses,"
   J. Phys. {\bf C}: Solid St. Phys., 13 (1980) L469-76

  \bibitem{Monasson}
  R.~Monasson, 
  ``Structural Glass Transition and the Entropy of the Metastable States,"
  {\it Phys. Rev. Let.}, Vol. {\bf 75}, Issue 15 (1995)
  [arXiv:cond-mat/9503166]

\bibitem{benawarner}
  I.~Bena, C.~-W.~Wang, N.~P.~Warner,
  ``Mergers and typical black hole microstates,''
  JHEP {\bf 0611}, 042 (2006).
  [hep-th/0608217].

\bibitem{benawarner2}
P. Berglund, E. G. Gimon and T. S. Levi, ``Supergravity microstates for BPS black holes and black rings," JHEP 0606 (2006) 007 [arXiv:hep-th/0505167].

\bibitem{benawarner3}
I. Bena and N. P. Warner, ``Bubbling supertubes and foaming black holes," Phys. Rev. D 74 (2006) 066001 [arXiv:hep-th/0505166].

\bibitem{benawarner4}
I. Bena, C. W. Wang and N. P. Warner, ``The foaming three-charge black hole," [arXiv:hep-th/0604110].

\bibitem{benawarner5}
V. Balasubramanian, E. G. Gimon and T. S. Levi, ``Four dimensional black hole microstates: From D-branes to spacetime foam," [arXiv:hep-th/0606118].

\bibitem{benawarner6} A. Saxena, G. Potvin, S. Giusto and A. W. Peet, ``Smooth geometries with four charges in four dimensions," JHEP 0604 (2006) 010 [arXiv:hep-th/0509214].

\bibitem{miranda} M.~C.~N.~Cheng,  
``More Bubbling Solutions,''
  JHEP {\bf 0703}, 070 (2007).
  [hep-th/0611156].

\bibitem{Hawking:1976de}
  S.~W.~Hawking,
  ``Black Holes and Thermodynamics,''
  Phys.\ Rev.\  {\bf D13}, 191-197 (1976).
  

\bibitem{Gross:1982cv}
  D.~J.~Gross, M.~J.~Perry, L.~G.~Yaffe,
  ``Instability of Flat Space at Finite Temperature,''
  Phys.\ Rev.\  {\bf D25}, 330-355 (1982).



\bibitem{Page:1981an}
  D.~N.~Page,
  ``Black Hole Formation In A Box,''
  Gen.\ Rel.\ Grav.\  {\bf 13}, 1117-1126 (1981).

\bibitem{Hawking:1982dh}
  S.~W.~Hawking, D.~N.~Page,
  ``Thermodynamics of Black Holes in anti-De Sitter Space,''
  Commun.\ Math.\ Phys.\  {\bf 87}, 577 (1983).

\bibitem{Chamblin:1999tk}
  A.~Chamblin, R.~Emparan, C.~V.~Johnson, R.~C.~Myers,
  ``Charged AdS black holes and catastrophic holography,''
  Phys.\ Rev.\  {\bf D60}, 064018 (1999).
  [hep-th/9902170].

\bibitem{Carlip:2003ne}
  S.~Carlip, S.~Vaidya,
  ``Phase transitions and critical behavior for charged black holes,''
  Class.\ Quant.\ Grav.\  {\bf 20}, 3827-3838 (2003).
  [gr-qc/0306054].

\bibitem{Bredberg:2010ky}
  I.~Bredberg, C.~Keeler, V.~Lysov, A.~Strominger,
  ``Wilsonian Approach to Fluid/Gravity Duality,''
  JHEP {\bf 1103}, 141 (2011).
  [arXiv:1006.1902 [hep-th]].



\bibitem{Bena:2011fc}
  I.~Bena, A.~Puhm, B.~Vercnocke,
  ``Metastable Supertubes and non-extremal Black Hole Microstates,''
  [arXiv:1109.5180 [hep-th]].

\bibitem{Chowdhury:2011qu}
  B.~D.~Chowdhury, B.~Vercnocke,
  ``New instability of non-extremal black holes: spitting out supertubes,''
 [arXiv:1110.5641 [hep-th]].
















  

  
 


    









%
%
%
%
%
%
%
%
%
%
%
%
%
%
%
%
%
%
%
%
%
%
%
%
%
%
%
%
%
%
%
%
%
%
%
%
%
%
%
%
%
%
%
%
%
%
%
%
%
%
%
%
%
%
%
%
%
%
%
%
%
%
%
%
%
%
%

\end{thebibliography}
\end{document}